\documentclass[journal]{IEEEtran}
\usepackage{cite,wrapfig}
\ifCLASSINFOpdf
  \usepackage{graphicx}
\else
  \usepackage{graphicx}
\fi
\usepackage{amsmath,color}
\usepackage{array}
\usepackage[caption=false, =normalsize,labelfont=sf,textfont=sf]{subfig}
\newcommand{\bmat}[1]{\ensuremath \begin{bmatrix}#1\end{bmatrix}}
\newcommand{\uu}{\ensuremath \boldsymbol{u}}
\newcommand{\pow}{\ensuremath \boldsymbol{p}}
\newcommand{\dd}{\ensuremath \boldsymbol{d}}
\newcommand{\ddh}{\ensuremath \boldsymbol{\hat{d}}}
\newcommand{\iload}{\ensuremath i_{\text{load}}^{\{d,q\}}}
\newcommand{\dq}{\ensuremath {\{d,q\}}}
\newcommand{\qd}{\ensuremath {\{q,d\}}}
\makeatletter
\newcommand{\xRightarrow}[2][]{\ext@arrow 0359\Rightarrowfill@{#1}{#2}}
\makeatother
\begin{document}

\title{Control Design for Inverters:\\ 
Beyond Steady-State Droop Laws}

\author{Alireza~Askarian\textsuperscript{1,a},~\IEEEmembership{Student Member,~IEEE,}
        Jaesang~Park\textsuperscript{1,b},~\IEEEmembership{Student Member,~IEEE,}
        and~Srinivasa~Salapaka\textsuperscript{1,c},~\IEEEmembership{Senior Member,~IEEE}
\thanks{\textsuperscript{1} Department of Mechanical Science and Engineering, University of Illinois at Urbana-Champaign, 61801 IL, USA}
\thanks{\textsuperscript{a}askaria2@illinois.edu, \textsuperscript{b}jaesang4@illinois.edu, \textsuperscript{c}salapaka@illinois.edu}
\thanks{The authors would like to acknowledge the support of Advanced Research Projects Agency-Energy (ARPA-E) for supporting this research through the project titled ARPA-E OPEN titled "Rapidly Viable Sustained Grid" via grant no. DE-AR0001016.}}

\maketitle

\begin{abstract}
This paper presents a novel control structure and control synthesis method for regulating the output voltage/frequency and power injection of DC-AC inverters. The traditional droop method offers attractive solution to achieve compromise between clashing power and voltage/frequency regulation objectives. However, it relies on use of nonlinear power variables through slow outer control loop. In this paper we formulate the traditional droop method as a feedback control problem based on static power-flow equations and show how neglecting the dynamics of inverter and transmission line restricts the attainable closed-loop bandwidth and stability and robustness margin. Then we introduce a mapping between power variables and current in $dq$ frame under given PLL condition, allowing for replacing the fast acting current variables as a proxy for power. Consequently, we present a novel control structure and control synthesis method based on disturbance rejection framework, and demonstrate inherent droop like characteristics in underlying dynamics for special cases of resistive and inductive line. Moreover, we generalize the proposed control synthesis procedure to include a generalized complex line dynamical model and introduce concept of hybrid-sourced-intverter. Finally, we validate higher bandwidth and better transient performance of our proposed design through experimental validation.
\end{abstract}


\IEEEpeerreviewmaketitle

\section{Introduction}

\IEEEPARstart{T}{he} existing power-grid system, which is rightly considered as one of the most outstanding achievements of the 20th century, has its limitations, many of which arise due to its top-down architecture and operation based on centralized power generation. Although the centralized generation has the advantage of less complexity in terms of system-wide voltage and frequency regulation, and stability, it does not provide the required flexibility for high-scale integration of renewables such as solar and wind turbines
\cite{inman2013solar}. These top-down, centralized networks cannot tolerate high margins of uncertainties and are vulnerable to single points of failure. Reliable and robust power distribution in these networks requires accurate estimates of power generation and consumption, which cannot be obtained with distributed and uncertain power sources. These shortcomings have led to significant interest in employing microgrids, which manage distributed generation sources and loads, and can robustly integrate renewables into the power network. These low-inertia systems respond fast to load and generation uncertainties enabled by high-bandwidth power electronics and digital signal processors, thus making smart-energy framework possible.\\
Nonetheless, stability and performance analysis of distributed systems such as microgrids are complicated. This complexity is typically addressed by imitating established power flow control methods for synchronous generators on microgrids. To this end, a set of voltage sourced inverters, operating in parallel, are modified to emulate the generator's drooping characteristic\cite{liu2015comparison}. The droop-control method's main advantages are easy implementation and intuition-guided analysis and design based on the analogy with well-studied generator sets\cite{chandorkar1993control}. Additionally, it provides a method to manage the net-load mismatch through a controlled deviation of voltage and frequency from their setpoints. Moreover, droop-based methods for microgrids enable plug-and-play capability, yield extra reliability and flexibility concerning uncertainties in communication-network (even with slow or no communication)\cite{chandorkar1996decentralized}, and enable decentralized architectures with parallel operation of multiple power sources; thereby avoid a single point of failure. 

However, simply mimicking generator dynamics on microgrids does not realize the full potential of power electronics. 
For example, high-bandwidth responses to uncertainties in loads and generation are sacrificed. 
Microgrids allow for fast responses to disturbances, which can be used to achieving better trade-offs between voltage and frequency regulation bandwidths, power-sharing accuracy, and robustness to load and generation uncertainties. Besides, the droop-based methods are primarily based on steady-state  power measurements, which are 
considerably slower than the time constants associated with the voltage/current dynamics, and result in lower   
Voltage and frequency regulation bandwidths\cite{golsorkhi2014control}. Furthermore, droop laws introduce an inherent trade-off between voltage regulation and power-sharing capability, \cite{guerrero2005output}, the sharing is negatively affected by line impedance dissimilarity in parallel operation\cite{vasquez2009adaptive} and they exhibit poor load harmonic sharing capability \cite{de2007voltage}.\\
This article develops a control system-theoretic framework that addresses these drawbacks. The control architecture directly uses fast voltage and frequency measurements of voltage sourced inverter for managing mismatches between demanded and available power. The resulting power flow is similar to the traditional droop-mechanisms. This framework retains the advantages of droop characteristics without requiring the slow active and reactive power calculation loops.\\
Here we present a control-system-based analysis of traditional droop design. We show how the steady-state power flow equation (in cases with purely resistive and inductive lines) and droop laws form a feedback system for output power regulation. We extend the steady-state power-based droop-control design to a general (not purely resistive or inductive) case. Here we exploit the algebraic structure of the underlying constitutive equations and derive droop parameters for a general output impedance case from parameters for a pure (say resistive) case, which guarantees identical voltage and frequency droop behavior to the pure case. Our analysis emphasizes the limitations of {\em constant-} parameter droop laws and motivates {\em dynamic} control designs.
An important feature of this framework is that control systems analysis and synthesis are based on {\em dynamic} (transfer function) models of the power systems components. It goes beyond steady-state considerations. 
The extra freedom in shaping closed-loop dynamics at different frequencies provides better tradeoffs between regulation, robustness, and power-sharing objectives. The control system provides robustness by compensating for the  {\em disturbances}-the uncertainties in loads and power generation, and {\em unmodeled} system dynamics. Furthermore, in our framework, we redefine input variables to control the system so that the constitutive dynamic equations in these variables are primarily linear, where uncertainties are modeled as {\em disturbance} terms. This enables employing widely available tools for linear system modeling, analysis, control synthesis, disturbance rejection, and implementation.\\
The closed-loop inverter system resulting from our control design can be interpreted as a hybrid-sourced-inverter (HSI); that is, it can be viewed as a voltage source interfaced to an ideal current source through a controllable impedance. This contrasts with popular voltage-source inverter (VSI) or control source inverter (CSI) designs applied to grid-isolated or grid-tied networks.   
We present control synthesis for general output impedance, which guarantees the same tradeoff in terms of robustness and performance as achieved for pure (resistive or inductive) output impedance cases.\\
The closed-loop dynamics preserve the steady-state behavior of droop designs while also achieve better robustness to uncertainties and higher-bandwidth regulation and power-sharing performance. Our experimental results demonstrate improvements in voltage regulation bandwidth (by factor of 2) compared to droop-based (steady-state oriented) design. Similarly, robustness margins are readily satisfied through considering full dynamics of line and inverter system.
\section{Review and Analysis of Droop Control}
In this section, we briefly review the principles of traditional droop control. Throughout this paper, for every signal $x(t)$, we denote its amplitude (in the corresponding phasor) by $\bar x$ and it's Laplace transform by   $\hat{x}(s)$. 

\begin{figure}
    \centering
    \includegraphics[width=0.9\linewidth]{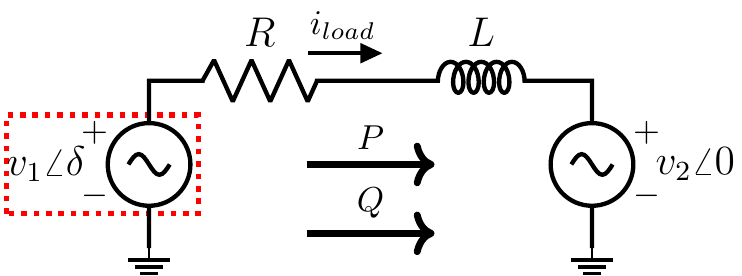}
    \caption{Traditional two voltage source system connected through complex impedance, used for power-flow analysis. The voltage source on the right side is considered as PCC reference voltage while the one on left represent voltage sourced inverter.}
    \label{fig: Two_Source}
\end{figure}

The main aim of droop control is to deliver demanded (active/reactive) power to an electrical node from another node, which is connected through a line impedance $\bar{Z}e^{j\phi}=R+jL\omega_{0}$ (See Fig. \ref{fig: Two_Source}). When there is a mismatch between the demanded power and actual power consumed, the droop control manages a controlled drop in voltage of a node.    For instance, In Fig. \ref{fig: Two_Source},  the nodes are represented by voltage sources, where  $\bar{v}_1\angle{\delta}$ and  $\bar{v}_2\angle{0}$ respectively represent the voltage phasors at the inverter-output (capacitor) powered by a distributed generator (DG) and the point of common coupling (PCC (grid)). The complex power flow from DG to PCC is given by \cite{de2007voltage}
\begin{equation*}
    \begin{split}
        P+jQ = \bar{v}_{1}e^{j\delta}
        \left( \frac{\bar{v}_{1}e^{j\delta}-\bar{v}_{2}}{\bar{Z}e^{j\phi}} \right)^{*},
    \end{split}
\end{equation*}
Therefore the 
the active and reactive power components are respectively given by
\begin{eqnarray}\label{eq: Comp_PF}
p:=\bmat{P\cr Q}=\bmat{\frac{\bar{v}_{1}^{2}}{\bar{Z}}\cos{(\phi)}-\frac{\bar{v}_{1}\bar{v}_{2}}{\bar{Z}}\cos{(\phi+\delta)}\cr \frac{\bar{v}_{1}^{2}}{\bar{Z}}\sin{(\phi)}-\frac{\bar{v}_{1}\bar{v}_{2}}{\bar{Z}}\sin{(\phi+\delta)}}.
\end{eqnarray}
Note that we can control $(\bar{v}_{1},\delta)$ through voltage-sourced inverter (VSI) in above equations. When we linearize (\ref{eq: Comp_PF}) with respect to $(\bar{v}_{1},\delta)$ around nominal operating point, $(\bar{v}_{2},0)$, we get
\begin{eqnarray}\label{eq: PF_Lin}
\pow:=\bmat{P\cr Q}= \frac{\bar{v}_{2}}{\bar{Z}}
\underbrace{\bmat{\cos(\phi)&\sin(\phi)\cr \sin(\phi)&-\cos(\phi)}}_{:=H_\phi}\bmat{\Delta \bar v\cr \bar v_2\delta}=:\rho_ZH_\phi\uu,
    \end{eqnarray}
where $\rho_Z=\bar{v}_2/\bar{Z}$, $\Delta v:=v_1-v_2$, and $\uu^T=[\Delta \bar{v}\ \ \bar{v}_2\delta]$. In high-voltage microgrids or when transformer or filter inductance is interfaced at the inverter's output terminals, the line impedance is assumed to be purely inductive. Under inductive line assumption we have $\phi \approx 90^{\circ}$ and $\bar{Z}=X$  reducing (\ref{eq: PF_Lin}) into $\pow=\rho_XH_{90^\circ}\uu$, where $H_{90^\circ}$ has $0$s as diagonal and $1$s in the anti-diagonal. This structure shows decoupling where the active power $P$ and reactive power $Q$ respectively depend strongly on phase difference $\delta$ and  the voltage difference $\Delta \bar v$. These dependencies motivate the familiar $P/f-Q/v$ droop design given by 
\begin{eqnarray}
        \bmat{
        \Delta \bar{v}\\\Dot{\delta}}=
       \underbrace{\bmat{0&k_{q}\cr k_{p}&0}}_{:=K_{90^\circ}}
        \bmat{P_{0}-P\\Q_{0}-Q}
        \label{eq: L_Droop}
\end{eqnarray}
where $\pow_0:=\{P_{0}, Q_{0}\}$    represents the nominal active and reactive power, and the anti-diagonal matrix  $K_{90^\circ}$ denotes the droop design parameters that decide how much steady-state actual voltage and frequency values deviate from the nominal values when the actual power $\pow$ deviates from the estimated setpoint $\pow_0$.
Similarly, for low-voltage microgrids where line impedance is mostly resistive, $\phi \approx 0$ and $\bar{Z}=R$, (\ref{eq: PF_Lin}) simplifies to decoupled equations $\pow=\rho_RH_{0^\circ}\uu$, where $H_0$ is a diagonal (identity) matrix. The corresponding  $P/v-Q/f$ droop design equations are given by
\begin{eqnarray}\bmat{\Delta v\cr \dot\delta}=\underbrace{\bmat{k_p&0\cr 0&k_q}}_{K_{0^\circ}}(\pow_0-\pow),
\label{eq: R_Droop}\end{eqnarray} where note that $K_{0^\circ}$ is a diagonal matrix. Droop equations shown above for the purely resistive and inductive cases are conventionally implemented by assuming that $\Delta \bar{v} \approx \bar{v}_{1} - v_{0}$ and $\Dot{\delta} \approx \omega_{1} - \omega_{0}$, where $\{v_{0},\omega_{0}\}$ represent nominal voltage and frequency at PCC\cite{de2007voltage}. In this case (\ref{eq: L_Droop}) can be rewritten as
{\small \begin{equation}
    \begin{split}
        \bar{v}_{1}(Q) &= v_{0} + k_{q} \left( Q_{0}-Q \right)\\
        \omega_{1}(P) & = \omega_{0} + k_{p} \left( P_{0}-P \right).
        \label{eq: L_Imp_Droop}
    \end{split}
\end{equation}}
Typical droop analysis assumes that the PCC voltage and frequency values $\bar{v}_{2}$ and $\omega_{2}$ are held constant at the nominal values and ignore the excursions from the nominal values. The effect of these excursions can be seen  by rewriting (\ref{eq: L_Imp_Droop}) as {\small \begin{equation}
    \begin{split}
        \begin{bmatrix}
        \Delta \bar{v}\\ \Dot{\delta}
        \end{bmatrix}=
        \begin{bmatrix}
        0&k_{q}\\
        k_{p}&0
        \end{bmatrix}
        \begin{bmatrix}
        P_{0}-P\\Q_{0}-Q
        \end{bmatrix}+
        \begin{bmatrix}
        v_{0}-\bar{v}_{2}\\ \omega_{0}-\omega_{2}
        \end{bmatrix},
    \end{split}
\end{equation}}
where we have used exact term for $\Delta \bar{v} = \bar{v}_{1}-\bar{v}_{2}$ and $\Dot{\delta}=\omega_{1}-\omega_{2}$. To evaluate the steady-state power-sharing performance under different loading scenarios in the system of parallel inverters, usually, the slope of $Q-v$ and $P-\omega$ plots are studied\cite{irving2000analysis}. Typical droop analysis ignores the disturbance term $[v_0-\bar v_2\ \omega_0-\omega_2]^\top$. In the next section, we show how this disturbance term is important in designing the droop coefficients and deciding the trade-off between the extent of the droop, power-sharing, and robustness to modeling uncertainties. We interpret the droop-control design as a feedback system and analyze the system to facilitate the droop design. Moreover, we consider an arbitrary (neither purely resistive nor inductive) line described by phasor $\bar{Z}\angle\phi$ and correspondingly generalized (coupled) droop laws. 
\section{Droop Design for General Output-Impedance Case }
\begin{figure}
    \centering
    \includegraphics[width=1\linewidth]{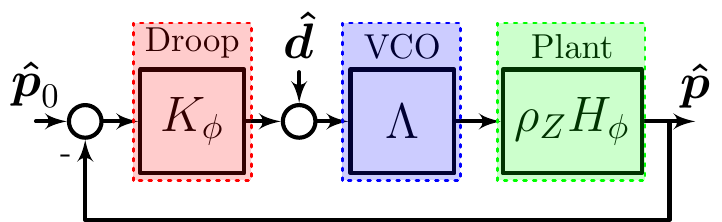}
    \caption{Control structure of traditional droop scheme. The plant and droop-law are diagonal matrix in case of resistive line, and anti-diagonal for inductive line. The VCO block comprise of matrix with integrator in second column of second row.}
    \label{fig: PQ_FB}
\end{figure}

Here we consider a generalized non-ideal voltage and frequency droop law in the Laplace domain as follows
\begin{eqnarray}
\bmat{\Delta\hat v\cr \widehat{\bar{v}_2\delta}}=\bmat{1&0\cr0&\frac{1}{s}}
\left(\bmat{k_{11}&k_{12}\cr k_{21}&k_{22}}\left(\hat\pow_0-\hat\pow\right)+\ddh\right),\label{eq: Gen_Droop}
\end{eqnarray}
where $\pow_0^{\top}=[P_{0}\enskip Q_{0}]$ and $\pow^{\top}=[P\enskip Q]$ denote reference and measured power and $\dd^\top=[d_v,d_w]^\top=[v_0-\bar v_2,\  \bar v_2(\omega_0-\omega_2)]^\top$ is the disturbance vector. Here the droop law can be interpreted as a compensator for power error signal in Fig. \ref{fig: PQ_FB}. Note that $K_{\phi}$ is a full matrix unlike purely resistive or inductive cases; thus results in coupled droop laws. Power-flow equation in (\ref{eq: PF_Lin}) along with generalized droop in (\ref{eq: Gen_Droop}) form a quasi-static closed-loop system shown in Fig. \ref{fig: PQ_FB}. After noting that $H_{\phi}=H_{\phi}^{\text{-}1}$ in equation (\ref{eq: PF_Lin}) and replacing $\boldsymbol{u}$ by the definition in (\ref{eq: Gen_Droop}) we can determine the closed-loop relationship from $\hat{\pow}_0$ and $\hat{\dd}$ to $\hat{\pow}$ as 
{\small \begin{equation}
    \begin{split}
        \hat{\pow}=\left( \rho_{Z}^{\text{-}1}H_{\phi}+\Lambda K_{\phi}\right)^{\text{-}1}
        \Lambda K_{\phi}\left(\hat\pow_0
        +K_{\phi}^{\text{-}1}\ddh\right).
        \label{eq: Dr_Fb}
    \end{split}
\end{equation}}
Designing $K_\phi$ to make $\pow$ track the desired $\pow_0$ is difficult since $K_\phi$ and $H_\phi$ are  full matrices. However, for decoupled systems, this design becomes simple; for instance, when $\phi=90^\circ$,  $H_{90^\circ}$ is anti-diagonal and with droop parameters $K_{90^\circ}$ in (\ref{eq: L_Imp_Droop}), equation (\ref{eq: Dr_Fb}) simplifies to 
{\small \begin{equation}
    \begin{split}
        \boldsymbol{\hat{p}}&=
       T_{90^\circ}\left(
        \boldsymbol{\hat{p}_{0}}+
        K_{90^\circ}^{\text{-}1}
        \boldsymbol{\hat{d}}\right),
        \label{eq: PQ_Loop}
    \end{split}
\end{equation}}
where $T_{90^\circ}= \begin{bmatrix}
        T_{P}&0\\
        0&T_{Q}
        \end{bmatrix}$ is a diagonal (decoupled) transfer function matrix given by 
{\small \begin{equation}
    \begin{split}
        T_{P} &= \frac{L_{P}}{1+L_{P}},\quad
        L_{P}=\frac{k_{p} \left(\bar{v}_{2}\right)^{2}}{Xs}\\
        T_{Q}&=\frac{L_{Q}}{1+L_{Q}},\quad
        L_{Q}=\frac{k_{q} \bar{v}_{2}}{X}.
        \label{eq: TP_Pow}
    \end{split}
\end{equation}}
Now we show that we can design $K_{\phi}$ for system with arbitrary $H_{\phi}$ such that feedback form in (\ref{eq: Dr_Fb}) mimics a system with desired $H_{\phi_{0}}$ and $K_{\phi_{0}}$ (e.g. $\phi_0=90^\circ$). Consider generalized droop matrix of following form
{\small \begin{equation}
    \begin{split}
        K_{\phi}=\frac{\rho_{Z_{0}}}{\rho_{Z}}\Lambda^{\text{-}1}
        R_{\Delta \phi}\Lambda K_{\phi_{0}},\quad
        R_{\Delta \phi}=
        \begin{bmatrix}
        \cos{\Delta \phi}&\sin{\Delta \phi}\\
        -\sin{\Delta \phi}&\cos{\Delta \phi}
        \end{bmatrix},
        \label{eq: DR_Rot}
    \end{split}
\end{equation}}
and $\Delta \phi = \phi - \phi_{0}$. Note that $R_{\Delta \phi}H_{\phi_{0}}=H_{\phi}$. Substituting $K_{\phi}$ from (\ref{eq: DR_Rot}) into (\ref{eq: Dr_Fb}), we get
\begin{eqnarray}
        \boldsymbol{\hat{p}}
        &=&\left(
        \rho_{Z_{0}}^{\text{-}1}H_{\phi_{0}}+
        \Lambda K_{\phi_{0}}
        \right)^{\text{-}1}
        \Lambda K_{\phi_{0}}\left(\boldsymbol{\hat{p}_{0}}
        +K_{\phi_{0}}^{\text{-}1}\boldsymbol{\hat{d}}\right),
        \label{eq: Dr_Rot}
    \end{eqnarray}
which depends only on $\phi_0$, and is identical to (\ref{eq: Dr_Fb}) with $\phi=\phi_0$. In case $\phi_0=90^\circ$, (\ref{eq: Dr_Rot}) is identical to (\ref{eq: PQ_Loop}). This simple control system manipulations result in multiple useful consequences.  From (\ref{eq: Dr_Rot}), it is evident that by designing droop for {\em a} line impedance value (say decoupled droop parameters for an inductive line where $\phi_0=90^\circ$), we can design for {\em any} line impedance by designing $K_{\phi}$ as in (\ref{eq: DR_Rot}). In fact choosing $\phi_0=90^\circ$,  the inductive line case, is preferred over the resistive case,  since the transfer function $T_P$ in (\ref{eq: PQ_Loop}) has an integrator ($1/s$) term, which implies zero steady-state error in tracking active power. Note that, our design assumes that $\phi=X/R$ ratio is known; if this ratio is not known, we can use line-impedance estimation methods to get an approximation of line impedance \cite{fantino2020grid,ciobotaru2007online,asiminoaei2005implementation}. Furthermore, any deviation from nominal values can be absorbed into disturbance term, which our control is designed to reject up to a specified bandwidth.
\begin{figure*}[!t]
    \centering
    \subfloat[]{\includegraphics[width=0.3\linewidth]{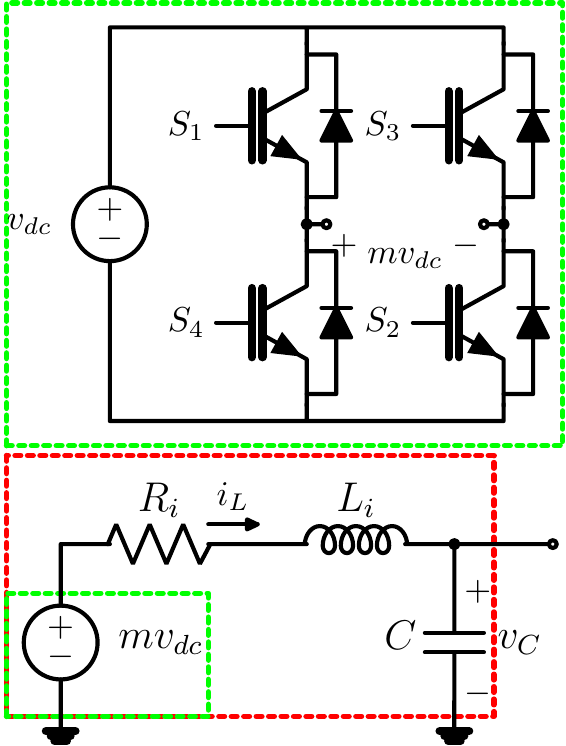}%
    \label{a}}
    \hfil
    \subfloat[]{\includegraphics[width=0.7\linewidth]{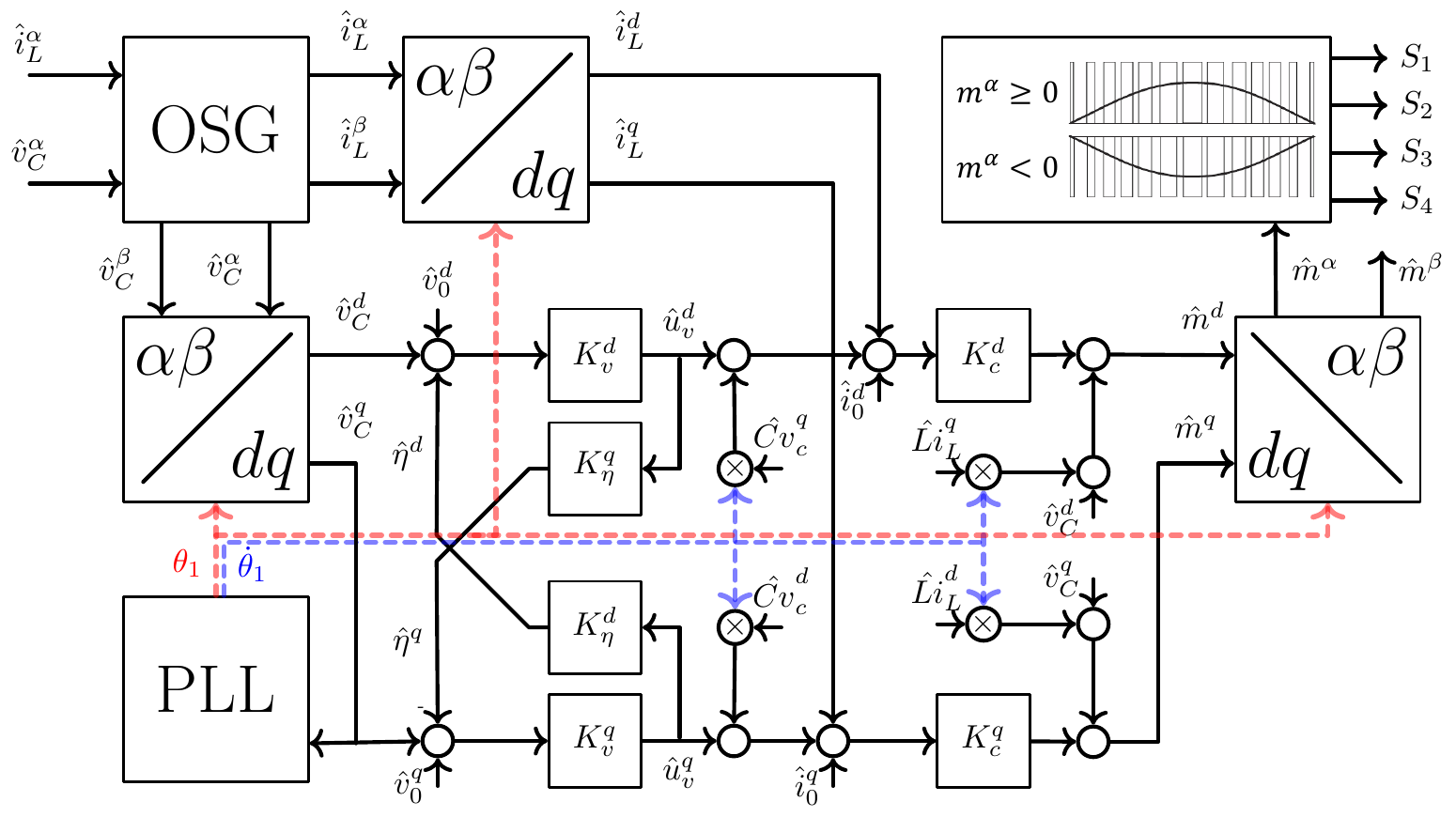}%
    \label{b}}
    \caption{(a) Switch node consisting of full-bridge connected to DC link with constant voltage $v_{dc}$. The half-bridge is used to modulate the signal $m$ using the pulse-width modulation (PWM) method. For high-frequency PWM, the switching node can be approximated by controlled voltage node $mv_{dc}$ interfaced with inverter output RLC filter. (b) For a single-phase inverter, the inductor current and capacitor voltage are treated as $\alpha$ components of the phasor, and the orthogonal part is generated using OSG block. The $\alpha\beta \to dq$ transform extracts the $dq$ components that are subsequently used as an input for embedded control algorithms. The output of the current compensator $u_{i}$ is used to calculate the $dq$ modulating signals based on (\ref{eq: m_index}). Finally, the $\{m^{d},m^{q}\}$ pair is transformed to $\{m^{\alpha},m^{\beta}\}$, where only $m^{\alpha}$ is used to calculate the gating signals.}
    \label{fig:InvAvg}
\end{figure*}
The choice of droop parameters is dictated by (\ref{eq: PQ_Loop}) and (\ref{eq: TP_Pow}).   To track the power variable coupled to voltage, $Q$ in (\ref{eq: PQ_Loop}), precisely we need to increase DC gain of $L_{Q}$, which is possible by either increasing $\bar{v}_{2}$, decreasing line impedance $X$ or increasing droop gain $k_{q}$. Usually, geographical parameters dictate the line impedance. However, we can improve the system power-sharing capabilities by increasing operating voltage and using high voltage (HV) designs. As a last resort, we can try to increase droop coefficient; this will cause performance degradation in voltage regulation under light or heavy loading condition as evident from droop equations (\ref{eq: L_Droop}) and (\ref{eq: R_Droop}) and violates small $\{\Delta \bar{v}, \delta\}$ assumption used to linearize power-flow equations in (\ref{eq: PF_Lin}).\\
One of the important outcomes of this analysis is that we can identify the limitations of the designs with constant droop parameters. From (\ref{eq: PQ_Loop}), it is clear that high values for droop parameters $\{k_{p},k_{q}\}$ will result in the system becoming less sensitive to disturbance $\ddh$ in PCC voltage and frequency. However, a high scalar droop coefficient can induce instability in the system; This instability is mainly due to unmodeled dynamics and transport delay. For example, including the digital signal processor (DSP) calculation transport delay $e^{\text{-}st_{0}}$, and neglected inverter voltage and frequency dynamics, $\{T_{v},T_{f}\}$, in droop equation (\ref{eq: L_Imp_Droop}), we get
{\small \begin{equation}
    \begin{split}
        L_{P}(s) &=
        \frac{k_{p} \left(\bar{v}_{2}\right)^{2}}{Xs}
        e^{\text{-}st_{0}}T_{f}(s),\quad
        \angle L_{P}(s)=\angle T_{f}(s)-90^{\circ}
        -st_{0}\\
        L_{Q}(s) &=
        \frac{k_{q} \bar{v}_{2}}{X}
        e^{\text{-}st_{0}}T_{v}(s),\quad
        \angle L_{Q}(s)=\angle T_{v}(s)
        -st_{0}.
        \label{eq: High_Dr}
    \end{split}
\end{equation}}
From (\ref{eq: High_Dr}), we can deduce that high values of droop parameters $\{k_{p},k_{q}\}$ results in reduced robustness (phase) margins. This conflict on the choice of droop parameters concerning performance and stability can be resolved by noting that we need high values of these parameters for good performance only near $0$ Hz, near the DC range. This motivates designing droop parameters as a transfer function instead of constant values as typically done. For instance, this dynamic droop design becomes very relevant  in the context of  
Droop operated inverters tied to weak grids, where high values for droop parameters are desired for better disturbance (voltage/frequency deviations) rejection. An approach is to include an integrator in $\{k_{p},k_{q}\}$ when operating in grid-tied mode, effectively turning the droop operated VSI into CSI with power reference. However, such a design is not robust since it results in two poles at origin and a $0$ phase margin. Considering that we have entirely neglected the inverter dynamics (described in the next section) and power flow dynamics in (\ref{eq: PF_Lin}), requiring model-robust outer power loop is a must; consequently, we can include additional lead compensator in $k_{p}$ leading to transfer function droop of the form
{\small \begin{equation}
	k_{p} = \frac{X}{\bar{v}_{2}^{2}}
    \left(
    \frac{s+\omega_{c}/\sqrt{a}}{s+\omega_{c}\sqrt{a}}
    \right)\frac{1}{s}, \quad
    L_{p} = \frac{1}{\sqrt{\alpha}}
    \left(
    \frac{s+\omega_{c}/\sqrt{a}}{s+\omega_{c}\sqrt{a}}
    \right)\frac{1}{s^2}
\end{equation}}
where $\omega_{c}$ is cross over frequency, and $a$ is a constant for adjusting the phase margin (see the lead-lag design \cite{skogestad2007multivariable} for details).
\section{VSI system and its dynamic model}
In microgrid systems, as stated earlier, the voltage $v_1$ in Fig. \ref{fig: Two_Source} is controlled using an inverter-based system, which comprises the full-bridge switch node shown in Fig. \ref{fig:InvAvg}(a), interfaced to the PCC through an output $RLC$ filter. This system takes power from a dc source at a constant voltage $v_{dc}$ and delivers appropriate demanded power across its output capacitor $v_C$; that is, it regulates the voltage amplitude $\bar v_C$ and frequency $\omega_1$ across the capacitor at constant values $v_0$ and $\omega_0$. The output power and voltage are controlled by designing the appropriate duty cycles of the switches in the inverter.

We obtain dynamic models for output capacitor voltage $v_C$ and inductor current $i_L$ that are averaged over a switching cycle to facilitate analysis \cite{yazdani2010voltage}; This averaging is justified since the switching frequency (on the order of $10$kHz) is much greater than the fundamental frequency ($60$Hz). This averaged model can be interpreted as a controlled voltage source with output voltage $m(t)v_{dc}$ where $m(t)\in [-1,1]$ is modulation index. The modulation index $m(t)$ is the control parameter, which can be actuated by appropriately designing  pulse-width modulated (PWM) signals which determine duty cycle $D_{S_k}$ (as fractions of switching periods) when each switch $S_k$ is closed; in fact for the bridge in Fig. \ref{fig:InvAvg}(a), the control signal $m(t)=D_{S1}-D_{S3}$ \cite{yazdani2010voltage}.

The equivalent circuit, as shown in Fig. \ref{fig:InvAvg}(a), represents averaged inverter dynamics with two state variables, capacitor voltage $v_{C}$ and inductor current $i_{L}$. Here we consider $i_{L}$ and $v_{C}$ as phasor quantities and utilize the following $\alpha-\beta$ and $d-q$ transform between variables in the fixed and rotating frame \cite{yazdani2010voltage}
\begin{eqnarray}
        v_{C}^{\alpha}+jv_{C}^{\beta}&=&
        (v_{C}^{d}+jv_{C}^{q})e^{j\theta_{1}(t)}\\
        i_{L}^{\alpha}+ji_{L}^{\beta}&=&
        (i_{L}^{d}+ji_{L}^{q})e^{j\theta_{1}(t)},
    \label{eq:dqab}
\end{eqnarray}
where $\theta_{1}(t)$ is the relative angle between the fixed and rotating frame, and superscripts $d-q$ and $\alpha-\beta$ denote orthogonal components of rotating and fixed frame. In single-phase inverters, the measured (real) voltage and current signals are treated as $\alpha$ component of phasor and the $\beta$ component is derived through orthogonal signal generator (OSG) as shown in Fig. \ref{fig:InvAvg}(b), and subsequently both components are fed into $\alpha\beta$ to $dq$ transform (see Fig. \ref{fig:InvAvg}(b)) to get corresponding $dq$ quantities \cite{yazdani2010voltage,golestan2012dynamics}. Here $\theta_{1}$ is obtained using a phased-lock loop (PLL) design presented in Section \uppercase\expandafter{\romannumeral 5}.A.(b).\\
The averaged dynamics of the inverter in the $d-q$ frame as shown in Fig. \ref{fig:InvAvg}(a) is given by\cite{yazdani2010voltage}
{\small \begin{equation}
\begin{split}
    L_{i}\frac{di_{L}^{d}}{dt} &= \underbrace{m^{d}(t)v_{dc} + L_{i}\Dot{\theta}_{1}(t)i_{L}^{q} - v_{C}^{d}}_{u_{i}^{d}(t)} - R_{i}i_{L}^{d},\\
    L_{i}\frac{di_{L}^{q}}{dt} &= \underbrace{m^{q}(t)v_{dc} - L_{i}\Dot{\theta}_{1}(t)i_{L}^{d} - v_{C}^{q}}_{u_{i}^{q}(t)} - R_{i}i_{L}^{q},\\
    C\frac{dv_{C}^{d}}{dt} &= i_{L}^{d} - i_{load}^{d} + C\Dot{\theta}_{1}(t)v_{C}^{q}, \\
    C\frac{dv_{C}^{q}}{dt} &= i_{L}^{q} - i_{load}^{q} - C\Dot{\theta}_{1}(t)v_{C}^{d},
    \label{eq: dq-eq}
\end{split}
\end{equation}}
where the dynamics of $d$ and $q$ components of $v_C$ and $i_L$ are {\em coupled}, and the underlying dynamics is {\em nonlinear}.  We use feedback linearization to decouple the inductor  current dynamics in (\ref{eq: dq-eq}). That is, here we use $u_{i}^{d}$ and $u_{i}^{q}$ as control design signals and the actuation parameters $\{m^{d},m^{q}\}$ (see Fig. \ref{fig:InvAvg}(b)) can easily be derived from them as
{\small \begin{equation}
    \begin{split}
        m^{d}(t)=\frac{u_{i}^{d}-L_{i}\Dot{\theta}_{1}i_{L}^{q}+v_{C}^{d}}{v_{dc}},\quad
        m^{q}(t)=\frac{u_{i}^{q}+L_{i}\Dot{\theta}_{1}i_{L}^{d}+v_{C}^{q}}{v_{dc}},
        \label{eq: m_index}
    \end{split}
\end{equation}}
where  $\{v_{C}^{d},v_{C}^{q}\}$, $\{i_{L}^{d},i_{L}^{q}\}$ and internal system signal $\Dot{\theta}_{1}$, $\{u_{i}^{d},u_{i}^{q}\}$ are obtained from real-time  measurements of 
voltage and current signals. From (\ref{eq: dq-eq}), it is clear that the resulting   dynamics for $d$ and $q$ components of $i_{L}$ are {\em linear} and decoupled in terms of the new control variables $u_{i}^{d}$ and $u_{i}^{q}$. In fact these dynamics are represented by identical transfer function given by
{\small \begin{equation}
    G_{c} =\frac{\hat{i}_{L}^{d}}{\hat{u}_{i}^{d}}
    =\frac{\hat{i}_{L}^{q}}{\hat{u}_{i}^{q}}
    =\frac{1}{L_{i}s + R_{i}}.
    \label{eq: Curr_Pl}
\end{equation}}
The physical constraint that $m(t)\in[-1\ 1]$ translates to saturation bounds on the control parameters $u_{i}^{\dq}$ which should respectively be within limits given by   
{\small
\[ \left[
        -v_{dc}-v_{C}^{\dq}+L_{i}\Dot{\theta}_{1}i_{L}^{\qd},\quad
        v_{dc}-v_{C}^{\dq}+L_{i}\Dot{\theta}_{1}i_{L}^{\qd}
        \right];\]}
\noindent therefore by having $v_{dc} \gg \bar{v}_{C}$ we can  avoid saturation related instability.\\
We address the coupling in capacitor voltage dynamics in (\ref{eq: dq-eq}), through feedback control, an appropriate structure to the control design, presented in the next section.
\begin{figure*}[!t]
    \centering
    \subfloat[]{\includegraphics[width=0.55\linewidth]{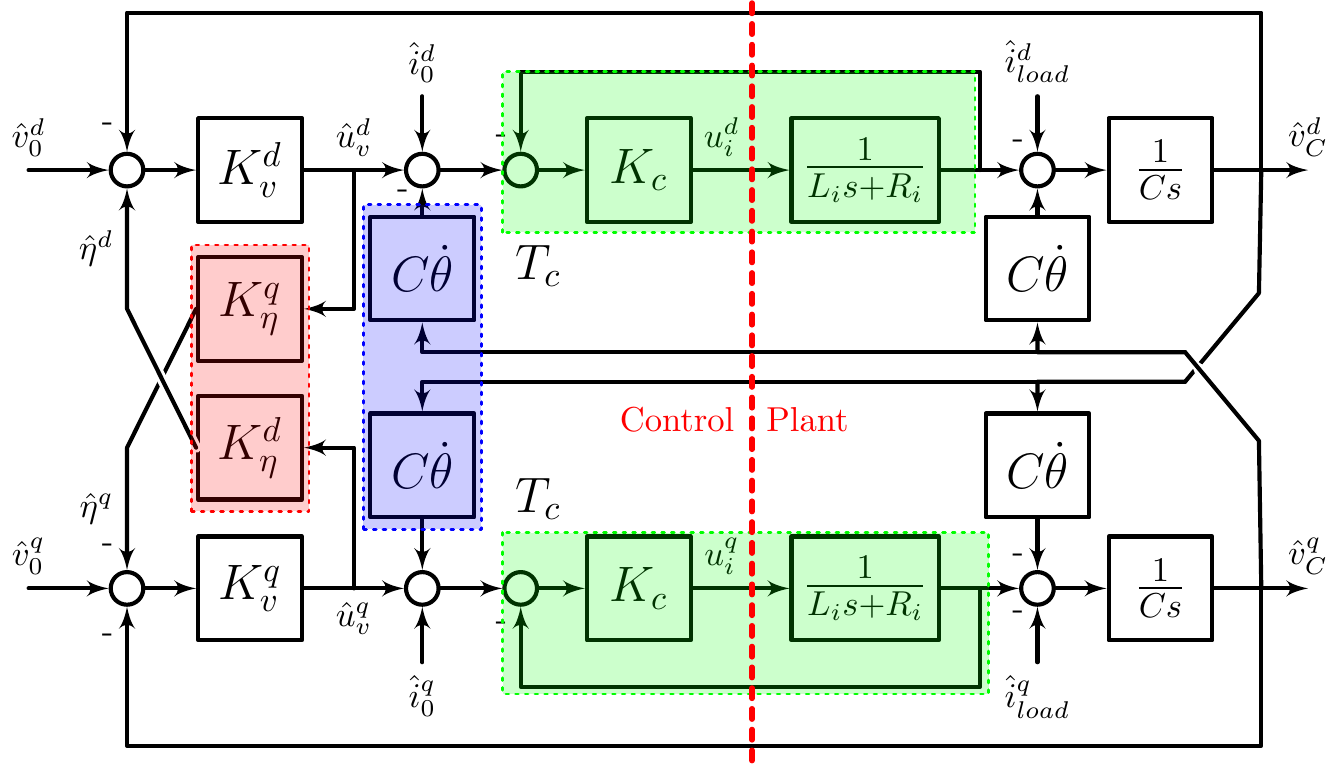}%
    \label{fig_first_case}}
    \hfil
    \subfloat[]{\includegraphics[width=0.45\linewidth]{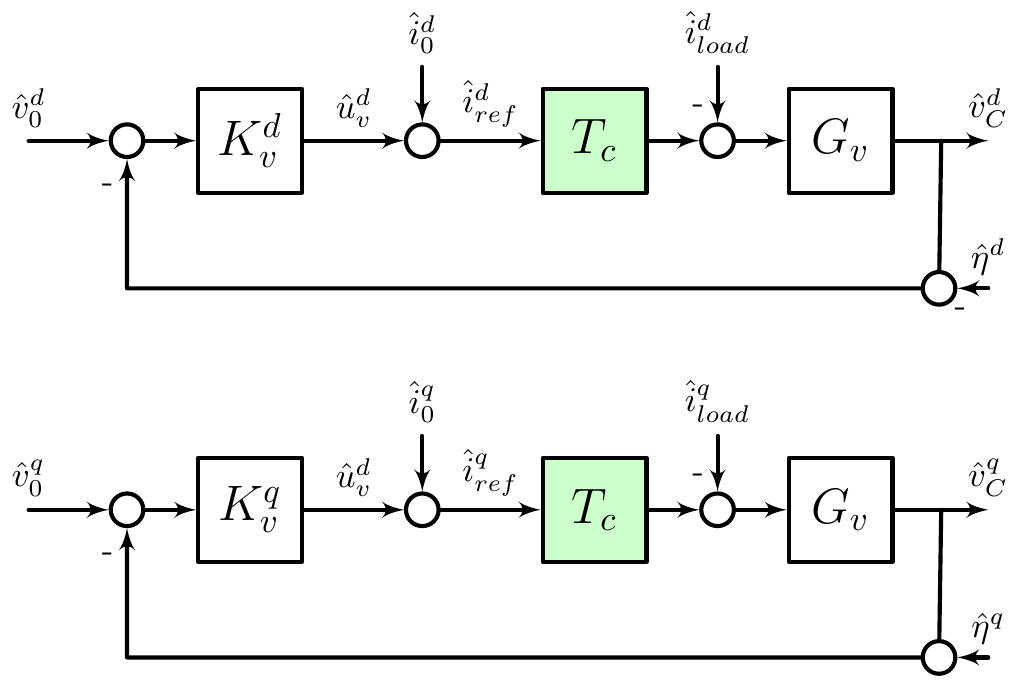}%
    \label{fig_second_case}}
    \caption{(a) Proposed control architecture for single VSI unit. The inner-current loop is marked with green box while the decoupling feedforward and coupling transfer functions are specified with blue and red boxes. (b) Equivalent control diagram after applying feedforward term. The effect of coupling transfer function is shown as exogenous input signals $\{\eta^{d},\eta^{q}\}$}
    \label{fig: Ctrl_Arch}
\end{figure*}
\section{Proposed Control Structure and Design}
\subsection{ \bf Control Objectives} The main control objectives are to guarantee (a) stability and robustness of the closed-loop system given the coupling terms in capacitor voltage dynamics and unknown load currents $\iload$ which we model as disturbances to our system, (b) performance requirements that include voltage regulation ($v_c^{\dq}\rightarrow v_0^{\dq}$), frequency regulation at the capacitor, and providing demanded active and reactive power when there is no mismatch between the demanded power and the actual consumed power; In case there is a mismatch, the control system is also required to manage the tradeoffs between the regulation and power tracking objectives, and (c) an inherent technical objective is estimating the rotating frame angle $\theta_1$ required for computation of $dq$ components of various signals.  

\subsubsection*{\bf \em  (a) Translation of objectives from power variables to load-current variables } In fact we translate the power objectives in terms of disturbance compensation corresponding to $\iload$, and show that our control design mimics droop-like power  management when $i_{\text{load}}$ deviates from demanded (designed) load values $i_0^{\dq}$. This is an important perspective proposed here, which enables system analysis in terms of linear system variables. The relation between Active $(P)$ and reactive power $(Q)$ and load current in $d-q$ frame
{\small \begin{equation}
\begin{split}
    \begin{bmatrix}
    P\\Q
    \end{bmatrix}
    =
    \begin{bmatrix}
    v_{C}^{d}& \hphantom{-}v_{C}^{q}\\
    v_{C}^{q}& -v_{C}^{d}
    \end{bmatrix}
    \begin{bmatrix}
    i_{load}^{d}\\
    i_{load}^{q}
    \end{bmatrix},
    \label{eq: pq-co}
\end{split}
\end{equation}}
is coupled based on the orientation of voltage phasor with respect to the rotating frame. The power is decoupled by fixing the rotating frame into voltage phasor, i.e., $v_{c}^{q}=0$, simplifying (\ref{eq: pq-co}) to
{\small \begin{equation}
    P = v_{C}^{d}i_{load}^{d}\quad \text{and} \quad Q = -v_{C}^{d}i_{load}^{q},
    \label{eq: Map_I_S}
\end{equation}}
which gives a direct mapping of current components $\{i_{L}^{d},i_{L}^{q}\}$ into active and reactive power with a coefficient equal to the magnitude of voltage. We set $v_0^d$ to be a {\em constant} value that we desire for the amplitude $v_C^d$ of the capacior voltage, and set $v_0^q=0 $ for fixing the voltage phasor into rotating frame. Regulation to these setpoints imply a near linear relationship between active/reactive powers $(P,Q)$ and load variables $(i_{\text{load}}^d,i_{\text{load}}^q)$. This  transformation enables our control design without implementing a slow outer power loop and corresponding calculations.

\subsubsection*{\bf \em  (b) PLL design objective for indirect control of phase-difference angle} The phase-difference variable $v_2\delta=v_2(\theta_1-\theta_2)$ is directly coupled to active/reactive power based on nature of transmission line as shown in (\ref{eq: PF_Lin}). In the context of our transformation of variables, this will equivalently relate the phase-difference to $i_{\text{load}}$. To get the PLL equation, we take the derivative of phase difference-variable
{\small \begin{equation}
    \begin{split}
        \frac{d\theta_{1}}{dt}=\dot{\delta}+\dot{\theta}_{2}.
    \end{split}
    \label{eq: PLL_PD}
\end{equation}}
Our objective is to control the phase-difference indirectly through the control of $\dot{\theta}_{1}$. However, the PCC frequency $\dot{\theta}_{2}$ is unknown; hence we treat it as a disturbance signal. Accordingly, we propose a PLL control law as
{\small \begin{equation}
    \frac{d\theta_{1}}{dt} = \omega_{1} = \omega_{0}
     + H(D)\frac{v_{C}^{q}}{\bar{v}_{2}},
    \label{eq: Res_PLL}
\end{equation}}
where $D=\frac{d(\cdot)}{dt}$ is  the  differential  operator and $\omega_{0}$ is the nominal frequency of PCC frequency $\dot{\theta}_{2}$. Replacing the PLL equation (\ref{eq: Res_PLL}) into phase-difference dynamics (\ref{eq: PLL_PD}) we get
{\small \begin{equation}
    \begin{split}
        \bar{v}_{2}\dot{\delta} = 
        \left(\omega_{0} - \dot{\theta}_{2}\right)\bar{v}_{2}
        +H(D)v_{C}^{q},
    \end{split}
    \label{eq: PLL_PD2}
\end{equation}}
where $(\omega_{0}-\dot{\theta}_{2})$ shows up as the disturbance signal into the system. Moreover, since we do not have direct measurements of $i_{\text{load}}$, we use quadrature component of inverter voltage $v_{C}^{q}$ instead of $i_{\text{load}}-i_0$ in equation (\ref{eq: PLL_PD2}). In the following section we show that $v_{C}^{q}$ embeds information for estimating load current.\\
Finally, observe that PLL (\ref{eq: Res_PLL}) should just be handled as part of the feedback path shown in equation (\ref{eq: PLL_PD2}) for regulating the $i_{load}$ to reference current.
\subsection{\bf Control Structure and Design}
The control design takes in the measurements $i_L$ and $v_C$ (and along with the internally generated signal $\dot\theta_1$ from PLL - see Fig. \ref{fig:InvAvg}(b)) to compute the control laws $u_i^d$ and $u_i^q$. The resulting control laws $u_i^d$ and $u_i^q$ are translated to modulation pair $\{m^{d}(t),m^{q}(t)\}$ via equations (\ref{eq: m_index}) and subsequently the $m^{\alpha}(t)$ is calculated using $dq\rightarrow \alpha\beta$ transformations as in (\ref{eq:dqab}). 

Fig. \ref{fig: Ctrl_Arch}(a) shows the inner-current outer-voltage control structure augmented with the inverter plant. This cascaded structure allows to inject the measurements $C\dot\theta v_c^{\dq}$ and nominal load signals $i_0^{\dq}$ in between the current and the voltage controllers. We use these injections and the design of the inner-loop controller $K_c$ to effectively decouple the voltage dynamics and diminish the effect of load disturbances. The decoupled outer-voltage loops are shown in Fig. \ref{fig: Ctrl_Arch}(b). The outer-voltage loop controllers $K_v^{\dq}$ is designed for voltage regulation and disturbance rejection for the entire loop. We elaborate on these designs below. 

\subsubsection*{\bf (a) Inner-Current Loop}
The inner-current loop, for both $d$ and $q$ axis, consists of simple first order plant $G_{c}$ in (\ref{eq: Curr_Pl}) and a compensator $K_{c}$ in series (Fig. \ref{fig: Ctrl_Arch}(a)), leading to closed-loop dynamics given by the transfer function
\begin{eqnarray}\label{eq:TC}
    T_{c}(s)=\frac{\hat{i}_{L}^{\dq}}{\hat{i}_{ref}}=\frac{G_{c}(s)K_{c}(s)}{1+G_{c}(s)K_{c}(s)}.
\end{eqnarray}
Note that from the resulting system in Fig. \ref{fig: Ctrl_Arch}(a), if we design $T_c(s)\equiv 1$, the coupling terms $C\dot\theta v_c^{\dq}$ in the plant (VSI) dynamics (\ref{eq: dq-eq}) will be canceled by our injected measurements in our control system. Also, the effect of load disturbance $\iload$ reduces to $\iload-i_0^{\dq}$, which becomes zero when there are no mismatch between the actual and nominal (demanded) load values. Accordingly, we design  $K_{c}$ as
\begin{eqnarray}
        K_{c}=\frac{1}{\tau}\left(L_{i} + \frac{R_{i}}{s} \right)\ \xRightarrow{\text{from}(\ref{eq:TC})}  T_{c}(s)=\frac{1}{\tau  s+1}.
        \label{eq: inner_comp}
\end{eqnarray}
This design achieves $T_c\approx 1$ and $S_c=1-T_c\approx 0$ over the bandwidth $[0,\ \tau^{-1}]$; furthermore, any sensor noise above the cutoff frequency $\tau^{-1}$ is attenuated. Moreover, the open loop transfer function is $L_{c}(s)=1/(\tau s)$, guaranteeing the stability and robustness of inductor current dynamics. Decoupling effect of coupling terms   $(C\dot{\theta}_{1})v_{C}^{\qd}$ and load disturbances $\iload$ becomes evident by writing the capacitor voltage equations given by $ \hat{v}_{C}^{\dq}=$
\begin{eqnarray}
       G_{v}\left(T_{c}\left(\hat{u}_{v}^{\dq}+i_{0}^{\dq}-T_c^{-1}\iload\right) + S_{c}
       \widehat{C\dot{\theta}_{1}v_{C}^{\qd}} \right),
\end{eqnarray}
where $G_v=1/Cs$. Here the effect of coupling terms is substantially attenuated over the bandwidth since $S_c\approx 0$, and the voltages depend only on load mismatches  $i_{0}^{\dq}-\iload$ rather than the load disturbances themselves.

\subsubsection*{\bf (b) Outer-Voltage Loop}
The voltage loop is augmented on top of the inner-current loop and regulates the capacitor voltage to the specified setpoint. Here we design control parameters $K_v^{\dq}$ that act on direct measurements $v_c^{\dq}$ and parameters $K_\eta^{\dq}$ that act on cross measurements $v_c^{\qd}$ (see Fig. \ref{fig: Ctrl_Arch}).  One of the distinguishing features of our control design is including the effects of the cross (coupling) terms as $\eta^{\dq}$ in each loop as shown in Fig. \ref{fig:InvAvg}(b) and \ref{fig: Ctrl_Arch}(a). These  coupling terms are given by $\hat{\eta}^{\dq}=K_{\eta}^{\dq}\hat{u}_{v}^{\qd}$, which compute to
{\small \begin{equation}
    \begin{split}
        \hat{\eta}^{d}(s)&=
        K_{\eta}^{d} \left( S^{q}K_{v}^{q}\hat{\eta}^{q}
        -T^{q}
        \left( \hat{i}_{0}^{q} - \frac{\hat{i}_{load}^{q}}{T_{c}(s)} \right) \right),\\
        \hat{\eta}^{q}(s)&=
        -K_{\eta}^{q}\left( S^{d}K_{v}^{d}\left( \hat{v}_{0}+\hat{\eta}^{d} \right)
        -T^{d}
        \left(\hat{i}_{0}^{d} - \frac{\hat{i}_{load}^{d}}{T_{c}(s)} \right) \right).
        \label{eq: Eta_Expand}
    \end{split}
\end{equation}}
The main purpose of the above design, is to include effect of $\left(\hat{i}_{0}^{d}-T_{c}^{\text{-}1\hat{i}_{load}^{d}}\right)$ into capacitor voltage dynamics (\ref{eq: quad_tf}) and $\left(\hat{i}_{0}^{q}-T_{c}^{\text{-}1\hat{i}_{load}^{q}}\right)$ into quadrature voltage dynamics. In section \uppercase\expandafter{\romannumeral 7}.B and \uppercase\expandafter{\romannumeral 7}.C we show how the coupling terms facilitate feedback control design in complex line scenario, and finally in (\ref{eq: coup_tf}), we present one such design.

With this control architecture, the  closed-loop voltages are given by $\hat{v}_{C}^{\dq}=$
{\small  \begin{equation}
    \begin{split}
        T^{\dq}\left(
        \hat{v}_{0}^{\dq}+\hat{\eta}^{\dq}(s)+\frac{1}{K_{v}^{\dq}(s)}
        \left(\hat{i}_{0}^{\dq}-\frac{\hat{i}_{load}^{\dq}}{T_{c}(s)}\right)
        \right),
        \label{eq: quad_tf}
    \end{split}
\end{equation}}
\noindent where 
   $     T^{\dq}=\frac{K_{v}^{\dq}T_{c}G_{v}}{1+K_{v}^{\dq}T_{c}G_{v}}.$
 From (\ref{eq: quad_tf}), it is evident that designing high-gain controllers $K_{v}^{\dq}$ such that both transfer functions $T^{\dq}\approx 1$ over a bandwidth will yield $v_C^{\dq}\approx v_0^{\dq}$. Also, high gain controllers will suitably attenuate the effect of load mismatch. However, we would like to design our controllers such that in the case of load mismatch, the voltage droops in a controlled manner.  We first present our control designs separately for purely resistive and inductive lines, and in section \uppercase\expandafter{\romannumeral 7}.C, we develop a design for a generic line impedance.

{\bf \em (a) Design for Resistive Line:}
For the $d$-loop regulation objective ($v_C^d\rightarrow v_0^d$), we propose a simple lag compensator
{\small \begin{equation}
    K_{v}^{d} = k\frac{s+z}{s+\beta^{d} z},
    \label{eq: Res_V_Comp}
\end{equation}}
where $\beta^{d} \ll 1$. This results in the loop transfer function in Fig. \ref{fig: Ctrl_Arch}(b) given by
{\small \begin{equation}
    L^{d} =G_vT_cK_{vd}= \frac{k}{C \tau} \left( \frac{s+z}{s+\tau^{-1}} \right)\frac{1}{s(s+\beta z)}.
\end{equation}}
Note that for such a system designing  $z \leq \tau^{-1}$ results in a large phase compensation given by 
\begin{align}
    \delta_{m} = \sin^{\text{-}1}\left( \frac{1-\tau z}{1+\tau z} \right)\  \text{at frequency}\ \omega_{m} = \sqrt{z\tau^{\text{-}1}}. 
    \label{eq: phase_comp}
\end{align}
Considering that $\beta$ and crossover frequency, $\omega_{c}$, are user-specified parameters, required phase compensation for at $\omega_{c}$ for stable and robust performance can be evaluated. Moreover, if we require $\omega_{m}=\omega_{c}$ then $\tau$ and $z$ are uniquely specified by (\ref{eq: phase_comp}). As last step, static gain $k$ is chosen such that $\left| L^{d}(j\omega_{c}) \right|=1$. Moreover, presence of integrator in the open loop guarantees DC reference tracking and disturbance rejection.

The $q$-loop regulation objective ($v_C^q\rightarrow v_0^q=0$), which  effectively decouples equation (\ref{eq: pq-co}) into (\ref{eq: Map_I_S}),   requires an effective disturbance rejection on $q$ axis. We propose a simple proportional-integrator (PI) compensator given by 
{\small \begin{equation}
    \begin{split}
        K_{v}^{q} = k\frac{s+z}{s}.
        \label{eq: Quad_V_Comp}
    \end{split}
\end{equation}}
With this design, the reactive load-current disturbance will be completely rejected at steady-state due to pure integrator. Here we design cross-coupling control transfer functions $K_{\eta}^{q}(j0)=0$, which implies that $\eta^{q}=0$, and therefore all three terms on the right-hand side of (\ref{eq: quad_tf}) go to zero in steady-state.

\subsubsection*{PLL Filter}
 The information on active and reactive load-current is already embedded in quadrature voltage through control architecture, as shown in equations (\ref{eq: quad_tf}) and (\ref{eq: Eta_Expand}). However, to propagate this information into $\bar{v}_{2}\delta$, we need to design a PLL filter $H(s)$ to recover, at least, the steady-state load-current information from quadrature voltage. In $q$-loop design, we proposed a disturbance rejection framework for $q$ voltage loop that lead to pole at origin in $K_{v}^{q}$ and zero at origin in $K_{\eta}^{q}$. More precisely  setting $\hat{v}_{0}^{q}=0$ in (\ref{eq: quad_tf}), we get 
 {\small \begin{equation}
    \begin{split}
        \widehat{
        \bar{v}_{2}\delta
        }=
        \frac{T^{q}H(s)}{s}
        \left(
        \eta^{q}(s)+\frac{1}{K_{v}^{q}(s)}
        \left(\hat{i}_{0}^{q}-\frac{\hat{i}_{load}^{q}}{T_{c}(s)}\right)
        \right)+\frac{\hat{d}_{\omega}}{s}.
        \label{eq: phase_dyn}
    \end{split}
\end{equation}}
Hence any PLL filter design, $H(s)$, with integrator cancels out the effect of the aforementioned pole and zero and recovers the load information (\ref{eq: phase_dyn}). Here we design a PLL filter with the following simple PI form 
{\small \begin{equation}
    H(s)=\frac{s+\beta^{q} z}{s}.
    \label{eq: PLL_tf}
\end{equation}}
 
{\bf \em (b) Design for Inductive Line:}
For inverters operating in the purely inductive microgrid, the voltage controller is a simple PI compensator
{\small \begin{equation}
    K_{v}^{d} = k\frac{s+z}{s}.
    \label{eq: Ind_V_Comp}
\end{equation}}
Note that PI compensator is a particular case of lag compensator given in (\ref{eq: Res_V_Comp}) where $\beta=0$. Therefore, we adopt the same methodology for specifying $\tau$ and $z$. However, it is noteworthy to consider two particular choices for phase margin, specifically $\delta_{m}=45^{\circ}$ and $\delta_{m}=53^{\circ}$\cite{yazdani2010voltage}. In the first case, the closed-loop will have a real pole at $-\omega_{c}$ and two complex conjugate poles with damping ratio $\xi = 0.707$. In the second case, closed-loop will have a triple pole at $-\omega_{c}$. In section \uppercase\expandafter{\romannumeral 7}.C, we explain why the specific choice of $\beta=0$ was adopted for the inductive line.

For the $q$-loop regulation objective, we employ an identical control design (\ref{eq: Quad_V_Comp}) as the resistive line case. Therefore reactive load-current disturbance will be completely rejected at steady-state due to pure integrator in voltage compensator. The only distinguishing issue is the presence of the coupling term $\eta^{q}$. Presence of pure pole in both $K_{v}^{q}$ and $G_{v}$ makes $\lim_{\omega \to 0}T^{q}(j\omega)=1$ and $\lim_{\omega \to 0}S^{q}(j\omega)=0$, therefore sufficient condition for full disturbance rejection on quadrature voltage is inclusion of blocking zero in coupling transfer function, $K_{\eta}^{q}(j0)=0$, making all three terms on right hand side of (\ref{eq: quad_tf}) go to zero in steady-state. Here we design  
{\small  \begin{equation}
    \begin{split}
        K_{\eta}^{d}=\frac{\alpha^{d}z}{s+z}, \quad
        K_{\eta}^{q}=\frac{\alpha^{q}z}{s + z}H(s)^{\text{-}1},
        \label{eq: coup_tf}
    \end{split}
\end{equation}}
where  PLL filter has the same simple PI form given by (\ref{eq: PLL_tf}). 
The choices made for $\{K_{\eta}^{d},K_{\eta}^{q}\}$ pair, and $H(s)$ in (\ref{eq: coup_tf}) and (\ref{eq: PLL_tf}) are not unique. However, it is one of the simplest form such that $H(s)\hat{v}_{C}^{q}(s)$ at steady-state, when $\boldsymbol{i}_{load}$ can be treated as constant, includes linear combination of $d-q$ current error given as
{\small  \begin{equation}
    H(j0)v_{C}^{q}=
    \alpha^{q}\left(i_{0}^{d}-i_{load}^{d}\right)
    +
    \frac{\beta^{q}}{k}
    \left(i_{0}^{q} - i_{load}^{q}\right).
\end{equation}}
indicating that a simple PI filter can recover steady-state load information from $v_{C}^{q}$.\\
With both inner, outer controllers and PLL specified, we are ready to present the initial outcome of adopting architecture in Fig. \ref{fig: Ctrl_Arch}(a) along with the controllers, as mentioned earlier.

\section{Analysis and Discussion}
\subsection{Droop Management} This section shows how the control structure and compensator design introduced in this paper leads to droop-like response at steady-state, eliminating the requirement for a slow outer power loop and additional stability issues connected to tweaking droop coefficients.
\subsubsection*{\bf\em (a) Resistive Control Structure}
With the control design described in the previous section, the closed-loop  $d-q$ component of capacitor voltage in Fig. \ref{fig: Ctrl_Arch}(a) are given by, 
{\small  \begin{equation}
    \begin{split}
        \hat{v}_{C}^{d} &= T^d(s) \bigg(\hat{v}_{0}^{d} + \frac{1}{K_{v}^{d}(s)} 
        \left(\hat{i}_{0}^{d} - T_{c}^{\text{-}1}(s) \hat{i}_{load}^{d} \right) \bigg),\\
        \hat{v}_{C}^{q} &= \frac{T^{q}(s)}{K_{v}^{q}(s)} 
        \left(\hat{i}_{0}^{q} - T_{c}^{\text{-}1}(s) \hat{i}_{load}^{q} \right).
        \label{eq: Res_In_Out}
    \end{split}
\end{equation}}
Equations (\ref{eq: Res_In_Out}) and (\ref{eq: Res_PLL}) at steady-state lead to the following static equations
{\small  \begin{equation}
    \begin{split}
        v_{C}^{d} &= v_{0}^{d} + \frac{\beta^{d}}{k}\left(i_{0}^{d} - i_{load}^{d} \right) ,\ 
        \omega_{1} = \omega_{0} + \frac{\beta^{q}}{v_{2}k}\left(i_{0}^{q} - i_{load}^{q} \right),
    \end{split}
\end{equation}}
which are identical to $i^{d}/v-i^{q}/f$ droop laws with coefficients $\{\beta^{d}/k,\beta^{q}/(v_{2}k)\}$ for voltage and frequency respectively.  
\subsubsection*{\bf \em (b) Inductive Control Structure}
Similarly for the inductive line case, the proposed control structure yields 
{\small \begin{equation}
    \begin{split}
        \hat{v}_{C}^{d} &= T^d(s) \bigg( \hat{v}_{0}^{d} + \hat{\eta}^{d}(s) + \frac{1}{K_{v}^{d}(s)} 
        \left(\hat{i}_{0}^{d} - T_{c}^{\text{-}1}(s) \hat{i}_{load}^{d} \right) \bigg),\\
        \hat{v}_{C}^{q} &= T^q(s) \bigg(\hat{\eta}^{q}(s) + \frac{1}{K_{v}^{q}(s)} 
        \left(\hat{i}_{0}^{q} - T_{c}^{\text{-}1}(s) \hat{i}_{load}^{q} \right) \bigg).
        \label{eq: Ind_In_Out}
    \end{split}
\end{equation}}
Adopting the same PLL equation as (\ref{eq: Res_PLL}) and considering the pure integrator in both $K_{v}^{d}$ and $G_{v}$, leads to following steady-state voltage and frequency equation
{\small  \begin{equation}
    \begin{split}
        v_{C}^{d} &= v_{0}^{d} - \alpha^{d}\left(i_{0}^{q} - i_{load}^{q} \right), \\
        \omega_{1} &= \omega_{0} + \frac{\alpha^{q}}{v_{2}}\left(i_{0}^{d} - i_{load}^{d} \right),
    \end{split}
\end{equation}}
that is current equivalent of $P/f-Q/v$ droop laws for inductive microgrids with coefficients $\{\alpha^{d},\alpha^{q}/v_{2}\}$.\\
\subsection{Interpretation of proposed control design as a Hybrid Source Inverter}
Inverters are commonly classified by their operating principle, namely current-sourced inverters (CSI) and voltage-sourced inverter (VSI) \cite{rocabert2012control}. CSIs, as the name suggests, operate as the current source, and they are ideal for situations where the third party, such as a grid, regulates voltage and frequency, and we need precise active and reactive power injection. 
\begin{figure}
\begin{center}
 \includegraphics[width=0.65\linewidth]{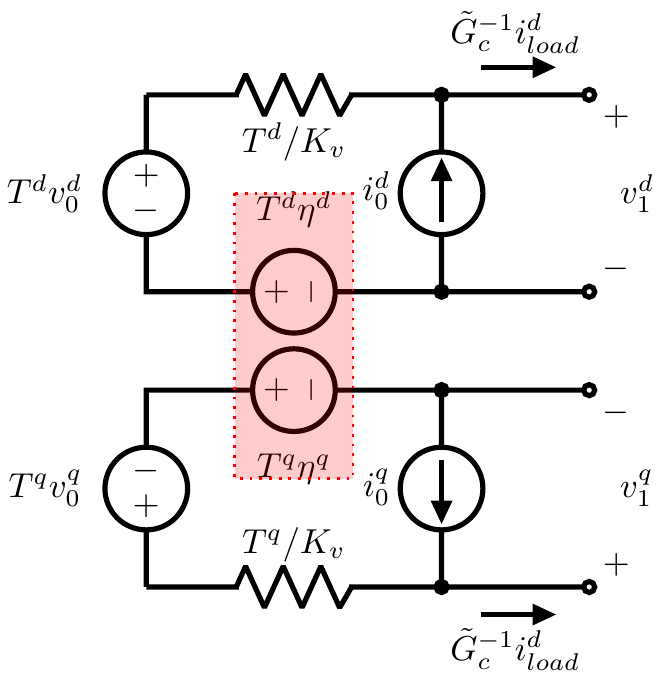}
    \caption{Circuit representation of hybrid source inverter, comprising of voltage source interfaced to an ideal current source through control induced impedance.}
    \label{fig: HSI}
    \end{center}
\end{figure}
On the other hand, VSIs are more suited for islanded operations where the network of inverters handles voltage, frequency, and power-sharing. However, the structure presented in this paper leads to the hybrid-sourced inverter (HSI), where the equivalent circuit that represents the control structure comprises a voltage source and current source in parallel (see Fig \ref{fig: HSI}). The HSI nature is based on form of equations that we derived in (\ref{eq: Res_In_Out}) and (\ref{eq: Ind_In_Out}), where the output voltage is function of both the reference voltage $\hat{v}_{0}^{d}$, and reference current $\{i_{0}^{d},i_{0}^{q}\}$. These equations can be regarded as KVL equation for the equivalent circuit in Fig.\ref{fig: HSI}, where the circuit parameters are given in terms of transfer functions to capture the frequency-wide dynamics of components; for example, the complementary sensitivity function represents the voltage and current source, attaining almost unity gain at designed bandwidth. Additionally, note that $K_{v}(s)$ as shown in (\ref{eq: Res_In_Out}) and (\ref{eq: Ind_In_Out}), represent voltage source equivalent series admittance and is used to achieve the balance between VSI or CSI nature of inverter. As we design $K_{v}$ such that $\lim_{\omega \to 0}|K_{v}(j\omega)|=\infty$, such as PI compensator, the output capacitor voltage more precisely follow the voltage reference. We move toward the VSI nature, on the other side as $\, lim_{\omega \to 0}|K_{v}(j\omega)|=0$ or removing the outer voltage compensator leads to CSI. The hybrid nature is specifically useful for designing a controller that can attain acceptable performance both in islanded and grid-tied operations.\\
Coupled voltage sources in the $d-q$ frame (see the box in Fig. \ref{fig: HSI}), is direct outcome inclusion of coupling terms $\{\eta^{d},\eta^{q}\}$, and mimics the steady-state function of the series inductor in $\alpha-\beta$ frame. We verify this by first considering the inductor voltage-current dynamic in the $d-q$ frame as
{\small  \begin{equation}
    \begin{split}
        \Delta v_{L}^{\dq} &= L\frac{di^{\dq}}{dt}-L\omega i^{\qd}, 
    \end{split}
\end{equation}}
and comparing it to (\ref{eq: Eta_Expand}) at steady-state 
{\small \begin{equation}
    \begin{split}
        \eta^{d}(j0) &= -K_{\eta}^{d}(j0)u_{v}^{q}
        =-K_{\eta}^{d}(j0)(i_{0}^{q}-i_{load}^{q}),\\
        \eta^{q}(j0) &= K_{\eta}^{q}(j0)u_{v}^{d}
        =K_{\eta}^{q}(j0)(i_{0}^{d}-i_{load}^{d}).
    \end{split}
\end{equation}}
If we assume that $K_{\eta}^{d}(j0)=K_{\eta}^{q}(j0)=L\omega$, then coupling terms $\eta^{d}$ and $\eta^{q}$ act as back emf of virtual inductor added to reference voltage. However, we can achieve more general results by exploiting asymmetrical degree of freedom in design of $\{K_{\eta}^{d},K_{\eta}^{q}\}$ pair.\\
In the next part, we express the proposed control design in the form of a feedback law to load the current $\boldsymbol{i}_{load}$ dynamical system.
\section{HSI Operation in Complex Line Scenario}
The traditional droop laws, as shown in (\ref{eq: L_Droop}) and (\ref{eq: R_Droop}), are based on a static power-flow equation in (\ref{eq: Comp_PF}) at the fundamental frequency. The static power-flow equation does not capture the dynamical and transient behavior of the system. We rectify this problem by considering $\boldsymbol{i}_{load}$ dynamics instead, which is directly related to injected power based on (\ref{eq: Map_I_S}).\\
In what follows, we derive the state-space model of transmission line in the form of an LTI system with control input and show how proposed HSI control design leads to natural feedback law to control injected load current and hence active and reactive power.
\subsection{Transmission Line as Controlled LTI System}
\begin{figure}
    \centering
    \includegraphics[width=0.9\linewidth]{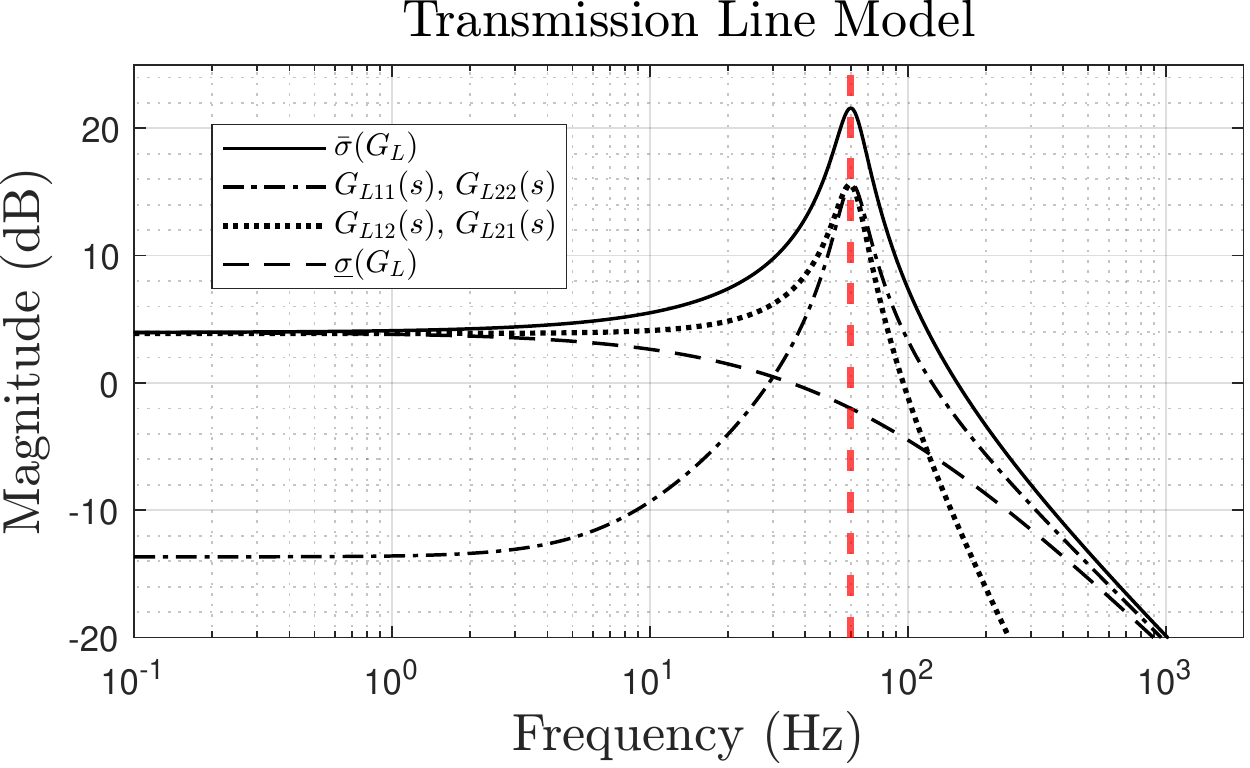}
    \caption{The magnitude of inductive line transfer function. The yellow plot signifies the anti-diagonal transfer functions while the red plot indicates the diagonal transfer functions. The vertical line signifies the transition boundary from the anti-diagonal to diagonal dominance.}
    \label{fig: Line_Model}
\end{figure}

The power flow equation in (\ref{eq: Comp_PF}) is obtained by evaluating impedance and voltage dynamics at grid fundamental frequency. However, by directly applying line dynamics, we can gain better insight into the power transaction's dynamic performance between inverter and PCC\cite{ferreira2019dynamic}. For this reason, we derive the dynamics of the line in Fig. \ref{fig: Two_Source}, in the $d-q$ frame as
{\small  \begin{equation}
    \begin{split}
        v_{1}^{\dq}-v_{2}^{\dq} &=L\frac{di_{load}^{\dq}}{dt} + i_{load}^{\dq}R
        - L\omega_{0} i_{load}^{\qd}
        \label{eq: Line_Dyn}
    \end{split}
\end{equation}}
Above derivation is independent of rotating frame orientation, however if we fix the frame on inverter voltage phasor, $v_{1}^{q}=0$, and use small angle approximation, $v_{2}^{d}=\bar{v}_{2}$ and $v_{2}^{q}=-\bar{v}_{2}\delta$, then above $d-q$ frame leads to $v_1^{d}-v_2^{d}\approx\bar{v}_1-\bar{v}_2=\Delta{\bar{v}}$ and $v_1^{q}-v_2^{q}\approx \bar{v}_{2}\delta$, turning (\ref{eq: Line_Dyn}) into
{\small \begin{equation}
    \begin{split}
        L\frac{d}{dt}
        \begin{bmatrix}
        i_{load}^{d} \\
        i_{load}^{q}
        \end{bmatrix}
        =
        \begin{bmatrix}
        -R & X \\
        -X & -R
        \end{bmatrix}
        \begin{bmatrix}
        i_{load}^{d} \\
        i_{load}^{q}
        \end{bmatrix}
        +
        \begin{bmatrix}
        \Delta \bar{v} \\
        \bar{v}_{2} \delta
        \end{bmatrix}
        \label{eq: Line_Dyn_Mat}
    \end{split}
\end{equation}}
where $X=L\omega_{0}$ represents line  reactance at fundamental frequency. Equation (\ref{eq: Line_Dyn}) depicts a LTI system of form $L\Dot{x}=Ax+Bu$, and power-flow equations given in (\ref{eq: L_Droop}) and (\ref{eq: R_Droop}) are special cases, evaluated at steady-state. Current-flow dynamics in (\ref{eq: Line_Dyn_Mat}) represents a coupled, second order, multiple-input multiple-output (MIMO) plant, $G_{L}$, with following transfer function representation\cite{zhang2009power}
{\small  \begin{equation}
    \begin{split}
        G_{L}
        =
        \frac{\boldsymbol{\hat{i}_{load}}}
        {\boldsymbol{\hat{u}}}
        =
        \frac{1/L^{2}}{\left( s^2 + 2\xi\omega_{n}s + \omega_{n}^{2}\right)}
        \begin{bmatrix}
        Ls+R & X \\
        -X & Ls+R
        \end{bmatrix}
        \label{eq: Line_TF}
    \end{split}
\end{equation}}
where natural frequency $\omega_n=\frac{\bar{Z}}{L}$, damping factor $\xi=\frac{R}{\bar{Z}}$, and $\bar{Z}=\sqrt{R^{2}+X^{2}}$ is magnitude of line impedance evaluated at fundamental frequency.\\
In case of complex line impedance, as shown in (\ref{eq: Line_TF}), both $d$ and $q$ component of load current, and consequently active and reactive power, depend on $\Delta \bar{v}$ and $\delta$. Fig. \ref{fig: Line_Model} shows the line's bode magnitude plot and corresponding singular values for the dominantly inductive line. The maximum and the minimum singular values come closer to each other a $\omega\to 0$ and $\omega\to \infty$; however, the difference between them is higher near the natural frequency $\omega\approx\omega_n$ (see Fig. \ref{fig: Line_Model}). Additionally, the natural frequency indicates the boundary where the plant transition from dominantly anti-diagonal to diagonal. This is evident by observing that for $\omega\to \infty$, the singular values converge to $\{G_{L11},G_{L22}\}$ while for $\omega \to 0$ it converges to $\{G_{L11},G_{L22}\}$.
\subsection{HSI as Feedback Law}
This section will use the proposed HSI architecture to derive the state feedback law to system in (\ref{eq: Line_TF}) and then do a meticulous analysis of the HSI unit's power-flow dynamics. If we rearrange dynamics of $v_{1}^{d}$ given in (\ref{eq: Ind_In_Out}) and phase equation (\ref{eq: phase_dyn}) assuming that $T^{d}=T^{q}\approx 1$ and $S^{d}=S^{q}\approx 0$, we get corresponding input control vector, $u^{\top} = [\Delta \bar{v}, \bar{v}_2 \delta]$ as
{\small  \begin{equation}
    \begin{split}
        \begin{bmatrix}
        \Delta \hat{v} \\
        \widehat{v_{2} \delta}
        \end{bmatrix}
        &=
        \Lambda
        \Bigg(
        \underbrace{
        \begin{bmatrix}
        \frac{1}{K_{v}^{d}} & -K_{\eta}^{d}\\
        HK_{\eta}^{q} & \frac{H}{K_{v}^{q}}
        \end{bmatrix}
        }_{K_{i}}
        \left(
        \boldsymbol{\hat{i}_{0}}-\boldsymbol{\hat{i}_{load}}
        \right)
        +
        \boldsymbol{\hat{d}}
        \Bigg).
        \label{eq: Inv_FB}
    \end{split}
\end{equation}}
The above equation indicates that by a smart selection of both control structure and disturbance rejection framework for a single inverter, we obtained naturally arising state-feedback law for current injection into PCC similar to the one achieved by generalized droop structure in (\ref{eq: Gen_Droop}), without the requirement for direct measurement of load-current and power calculation.\\
The complexity of this problem lies in four degree of freedom control structure, $\{K_{v}^{d},K_{v}^{q},K_{\eta}^{d},K_{\eta}^{q}\}$ and multi-input multi-output (MIMO) system. Like single-input single output (SISO) systems, the MIMO system's performance is studied through the design of sensitivity, $ S $, and complementary-sensitivity, $T$ transfer matrices. The closed-loop system is described by \begin{eqnarray}
        \hat{\boldsymbol{i}}_{load}&=&T\hat{\boldsymbol{i}}_{0}+SG_{L}\Lambda\hat{\boldsymbol{d}}
        \Rightarrow \hat{\boldsymbol{e}}=S\left(\hat{\boldsymbol{i}}_{0}+
        G_{L}\Lambda\hat{\boldsymbol{d}}\right)
\end{eqnarray}
where the sensitivity $S$ and complementary sensitivity $T$  transfer functions are given by {\small \begin{equation}
    \begin{split}
        S &= \bigg( I_{2} +
        G_{L}
        \Lambda
        K_{i}
        \bigg)^{\text{-}1},\quad
        T=I_{2}-S.
    \end{split}
    \label{eq: SVD_TS}
\end{equation}}
The requirement for command tracking and disturbance rejection is obtained by $T\approx1$ and $S\approx0$. However, in practical designs, sensitivity transfer function only remains small (less than $0.707$ or -3dB) within certain bandwidth, and considering.
{\small \begin{equation}
    \begin{split}
        \hat{\boldsymbol{e}}=S\left(\hat{\boldsymbol{i}}_{0}+
        G_{L}\Lambda\hat{\boldsymbol{d}}\right)
    \end{split}
\end{equation}}
The sensitivity bandwidth sets the frequency range where we have effective disturbance rejection and tracking performance\cite{skogestad2007multivariable} .\\
Unfortunately, unlike their SISO counterparts, loop-shaping of $ S $ and $T$ through compensator design is not straightforward since the MIMO system's performance, such as reference tracking and disturbance rejection, is directly linked to the loop-shape of the smallest and largest singular values of $S$ and $T$. For these reasons, modern loop-shaping methods, such as $\mathcal{H}_{\infty}$ synthesis \cite{skogestad2007multivariable}, They are well developed and suited for such MIMO problems. However, in what follows, we show that by setting specific constraints on matrix $K$, we can analyze steady-state tracking and disturbance rejection of the MIMO system.
\subsection{Tracking Performance}
\begin{figure}
    \subfloat[]{\includegraphics[width=0.53\linewidth]{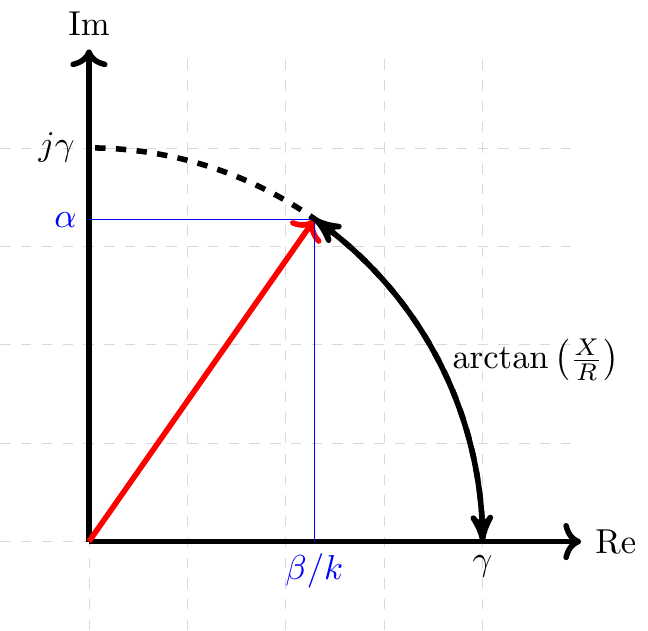}}
    \subfloat[]{\includegraphics[width=0.47\linewidth]
    {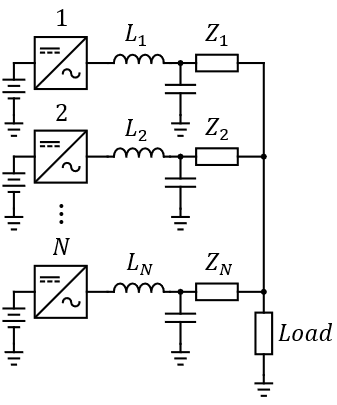}}
    \caption{(a) Geometric representation of scaled and normalized performance parameters $\beta/k$ and $\alpha$ and their relationship to complex line parameters $R$ and $L$. (b) A microgrid comprising of N parallel inverters interfaced to common load at PCC through the corresponding line impedances.}
    \label{fig: Droop_Cir}
\end{figure}

In this section we demonstrate tracking and disturbance rejection performance of MIMO line dynamics under specific design constraint on inverter compensators, $\{K_{v}^{d},K_{v}^{q}\}$ and $\{K_{\eta}^{d},K_{\eta}^{q}\}$. To this end, we use set of simplifying assumption, namely, $T^{d}=T^{q}=T_{c}^{\text{-}1} \approx 1$. Consider voltage compensators, $\{K_{v}^{d},K_{v}^{q}\}$ in (\ref{eq: Res_V_Comp}) and (\ref{eq: Quad_V_Comp}), coupling transfer function $\{K_{\eta}^{d},K_{\eta}^{q}\}$ and PLL filter $H(s)$ from (\ref{eq: coup_tf}) and (\ref{eq: PLL_tf});
Consequently the $K_{i}$ as given in (\ref{eq: Inv_FB}) turns into
{\small  \begin{equation}
    \begin{split}
    K_{i}
    =
    \frac{1}{k(s+z)}
        \begin{bmatrix}
        s+\beta^{d} z
        & 
        - k \alpha^{d} z\\
         k \alpha^{q} z
        &
        s+\beta^{q} z
        \end{bmatrix}.
        \label{eq: MIMO_Comp_tf}
    \end{split}
\end{equation}}
If we choose $\alpha$ and $\beta$ parameters in the following normalized and scaled form as shown in Fig. \ref{fig: Droop_Cir}(a)
{\small  \begin{equation}
    \begin{split}
        \frac{\beta^{d}}{\gamma^{d}k}=
        \frac{\beta^{q}}{\gamma^{q}k}=
        \frac{R}{\bar{Z}},
        \quad
        \frac{\alpha^{d}}{\gamma^{d}} =
        \frac{\alpha^{q}}{\gamma^{q}} =
        \frac{X}{\bar{Z}},
        \label{eq: Droop_Norm}
    \end{split}
\end{equation}}
then, at steady-state $K_{i}$ becomes 
{\small  \begin{equation}
    \begin{split}
        K_{i}(j0) = 
        \frac{1}{\bar{Z}}
        \begin{bmatrix}
        \gamma^{d} & 0\\
        0 & \gamma^{q}
        \end{bmatrix}
        G_{L}^{\text{-}1}(j0).
        \label{eq: MIMO_Comp}
    \end{split}
\end{equation}}

To evaluate the steady-state performance of the above controller, consider the equation (\ref{eq: SVD_TS}). Using MIMO compensator $K_{i}$ defined in (\ref{eq: MIMO_Comp}) we get
{\small  \begin{equation}
    \begin{split}
        S(j0)&= 
        G_{L}(j0)
        \begin{bmatrix}
        \frac{\bar{Z}}{\gamma^{d}+\bar{Z}}
        &0\\
        0&
        0
        \end{bmatrix}
        G_{L}^{\text{-}1}(j0)\\
        T(j0)&=
        G_{L}(j0)
        \begin{bmatrix}
        \frac{\gamma^{d}}{\gamma^{d}+\bar{Z}}
        &0\\
        0&
        1
        \end{bmatrix}
        G_{L}^{\text{-}1}(j0).
    \end{split}
    \label{eq: eig_sin}
\end{equation}}
Note that, $G_{L}(j0)^{\text{-}1}=\bar{Z}^{2}G_{L}(j0)^{\top}$, hence, both eigenvalues values in (\ref{eq: eig_sin}) are also singular values, of $S$ and $T$.\\
We assess steady-state tracking and disturbance rejection performance using smallest singular value of $T(j0)$, $\underline{\sigma}(T(j0))=\gamma^{d}/(\gamma^{d}+\bar{Z})$, which sets lower bound for worst case performance. On the other hand we have, $\overline{\sigma}(T(j0))=1$ showing that axis associated with second row of $T$ gets tracked with unity gain. This becomes more clear if we consider extreme cases of purely inductive line where $i_{load}^{d}$ is variable corresponding with second row, therefore it gets tracked more precisely. For disturbance rejection at steady-state we have, $\underline{\sigma}(S(j0)) = 0$ and $\overline{\sigma}(S(j0)) = |Z|/(\gamma^{d}+\bar{Z})$ indicating worst case disturbance rejection equivalent to $\overline{\sigma}(S(j0))$. Note that as scaling factor $\gamma^{d}$ increases, $\underline{\sigma}(T(j0))$ get closer to $1$ and the complementary sensitivity transfer matrix $T$ converges to identity matrix, achieving better active and reactive decoupling. This decoupling is direct result of high loop gain decoupling effect of feedback system proposed. Although, large scaling factor, $\{\gamma^{d},\gamma^{q}\}$, results in more accurate current tracking at steady-state, it can also cause large deviation in voltage and frequency. This is further clarified by considering steady-state operation of HSI equation (\ref{eq: Inv_FB}) with compensator (\ref{eq: MIMO_Comp_tf}) and parameters in (\ref{eq: Droop_Norm}) to get
{\small \begin{equation}
    \begin{split}
        v_{2}\left(\omega_{1}-\omega_{0}\right)&=
        \left(
        \alpha^{q}
        \Delta i^{d}
        +
        \frac{\beta^{q}}{k}
        \Delta i^{q}
        \right)
        =
        \gamma^{q} |\Delta i|\sin{(\phi+\phi_{\Delta i})}
        \\
        \left(v_{1}-v_{0}\right)&=
        \left(
        \frac{\beta^{d}}{k}
        \Delta i^{d}
        -
        \alpha^{d}
        \Delta i^{q}
        \right)=
        \gamma^{d} |\Delta i|\cos{(\phi+\phi_{\Delta i})}
        \label{eq: Dr_qual}
    \end{split}
\end{equation}}
where $\Delta i^{\top}=[i_{0}^{d}-i_{load}^{d}\enskip i_{0}^{q}-i_{load}^{q}]$, $\phi = \arctan(X/R)$ and $\phi_{\Delta i}=\arctan(\Delta i^{q}/\Delta i^{d})$. Furthermore, we can write down (\ref{eq: Dr_qual}) in more insightful form
{\small  \begin{equation}
    \begin{split}
        {|\Delta i|}=
        \sqrt{\left(\frac{v_{2}}{\gamma^{q}}\right)^{2}
        \left(\omega_{1}-\omega_{0}
        \right)^{2}+
        \left(\frac{1}{\gamma^{d}}\right)^{2}
        \left(v_{1}-v_{0}\right)^{2}},
    \end{split}
\end{equation}}
where right hand side indicates how voltage and frequency deviation are weighted through their corresponding scaling factor in case of load current deviation.\\
As concluding remark, we take a closer look at gains $\beta/k$ and $\alpha$. By design, explained earlier, we have $\beta/(k\alpha)=R/X$, therefore when $R/X \ll 1$ or in case of purely inductive line, $R=0$, we can choose $\beta=0$, this leads to PI design for $k_{v}$ and is the basis for the proposed design in section \uppercase\expandafter{\romannumeral 4}-B. On the other extreme, if we have $R/X \gg 1$ or purely resistive line, we can choose $\alpha=0$ or equivalently omitting coupling transfer functions from design as, again, shown in section \uppercase\expandafter{\romannumeral 5}.\textit{(b)}. 
\subsection{Disturbance Rejection In Grid-Tied Operation}
As shown in (\ref{eq: Inv_FB}), grid voltage and frequency deviation from nominal values show up as disturbances in the feedback system. This effect can be captured by writing (\ref{eq: Inv_FB}) in terms of $T$ and $S$ as
{\small \begin{equation}
    \boldsymbol{\hat{i}_{load}}
    =
    T\left(
    \boldsymbol{\hat{i}_{0}}
    +
    K_{i}^{\text{-}1}
    \boldsymbol{\hat{d}}\right)
\end{equation}}
We compute $K_{i}^{\text{-}1}$ using (\ref{eq: MIMO_Comp_tf}) and (\ref{eq: Droop_Norm}) as
{\small \begin{equation}
    \begin{split}
    K_{i}^{\text{-}1}&=\frac{k(s+z)}
    {s^{2}+2\xi \omega_{i}s
    +\omega_{i}^{2}}
        \begin{bmatrix}
        s+\beta^{q} z
        & 
         k \alpha^{d} z\\
        - k \alpha^{q} z
        &
        s+\beta^{d} z
        \end{bmatrix},
    \end{split}
\end{equation}}
where
{\small \begin{equation}
    \begin{split}
        \omega_{i}&=kz\sqrt{\gamma^{d}\gamma^{q}},
        \quad
        \xi_{i}=\frac{\gamma^{d}+\gamma^{q}}
        {2\sqrt{\gamma^{d}\gamma^{q}}}\frac{R}{|Z|}.
    \end{split}
\end{equation}}
In the above, line dynamics under the dominantly inductive grid, $R\approx0$, includes resonance frequency at $\omega_{i}$ with damping $\xi_{i}$ close to zero. We can circumvent the undesirable amplification of disturbance signals at $\omega_{i}$ by either selecting scaling factors to increase $\omega_{i}$ such that it lies out of desired tracking bandwidth or in a more elegant approach; we can augment band stop filter such as
{\small \begin{equation}
    \begin{split}
        H_{n}=
        \frac{s^{2}+2\xi\omega_{i}s+\omega_{i}^{2}}
        {s^{2}+2\xi_{0}\omega_{n}s+\omega_{i}^{2}},\quad
        \xi_{i} \ll \xi_{0}
        \label{eq: notch_fil}
    \end{split}
\end{equation}}
in control matrix $K_{i}$ to dampen the response of line at the resonance frequency. The augmented controller $H_{n}K_{i}$ is easily implemented by directly including the notch filter transfer function (\ref{eq: notch_fil}) into coupling transfer functions $\{H_{n}K_{\eta}^{d},H_{n}K_{\eta}^{q}\}$ and PR compensator of form 
{\small \begin{equation}
    \begin{split}
        H_{\text{PR}} = H_{n}^{\text{-}1}=1+
        \frac{2\left(\xi_{0}-\xi\right)\omega_{n}s}
        {s^{2}+2\xi\omega_{n}s+\omega_{n}^{2}}
    \end{split}
\end{equation}}
into voltage compensator pair $\{H_{\text{PR}}K_{v}^{d},H_{\text{PR}}K_{v}^{q}\}$.\\
We could have enhanced the current situation by adopting higher-order controllers, however without using advanced methods such as $\mathcal{H}_{\infty}$, designing the controller will become intractable.
\section{Experimental Validation}
In this section, we demonstrate the performance and validate the effectiveness of the proposed control structure. 
\begin{table}
\caption{Inverter Parameters}
\begin{center}
\begin{tabular}{||c c c c||}
\hline
& Inverter 1 & Inverter 2 & Inverter 3
\\[0.5ex]
\hline\hline 
     $C$ & 40$\mu \text{F}$  & 45$\mu \text{F}$& 35$\mu \text{F}$\\
     \hline
     $L_{i}$ & 3.3mH & 2.7mH & 3mH \\
     \hline
     $R_{i}$ & 0.2$\Omega$ & 0.2$\Omega$ & 0.2$\Omega$\\
     \hline
     $Z$ & $0.1+j0.7\Omega$ & $0.1+j0.75\Omega$ 
     & $0.1+j0.8\Omega$ \\
    \hline
    $\bar{v}_{dc}$ & 250 V& 250 V& 250 V\\
    \hline
    Sharing & 0.2 & 0.3 & 0.5\\
    \hline
\end{tabular}
\end{center}
\label{table: param}
\vspace{-4mm}
\end{table}
\subsection{Microgrid Topology And Parameters}
We consider a microgrid system comprising parallel inverters interfaced to PCC through their corresponding line impedance (see Fig. \ref{fig: Droop_Cir}(b)). The parallel topology is of particular interest since it can scale the power rating of microgrids by adding additional power sources and avoiding single points of failure through distributed power generation. However, the distributed nature of microgrids adds an extra level of complexity since the distributed system should collectively satisfy both the stability and performance bounds. Furthermore, we consider total communication blackout; each inverter does not share nor receive any type of information from other inverters. This capability is essential to assure both network resilience to communication failure and provide plug and play capability for power sources connected to the system via an inverter.\\
In our test case, we implement a high-fidelity model of three parallel switched inverters with design parameters given in Table \ref{table: param}. using OPAL-RT eFPGASIM. eFPGASIM is an FPGA-based hardware-in-the-loop (HIL) platform from OPAL-RT for high-fidelity sub-microsecond real-time simulation of circuits.\\
\begin{figure}
  \begin{center}
  \includegraphics[width=0.9\linewidth]{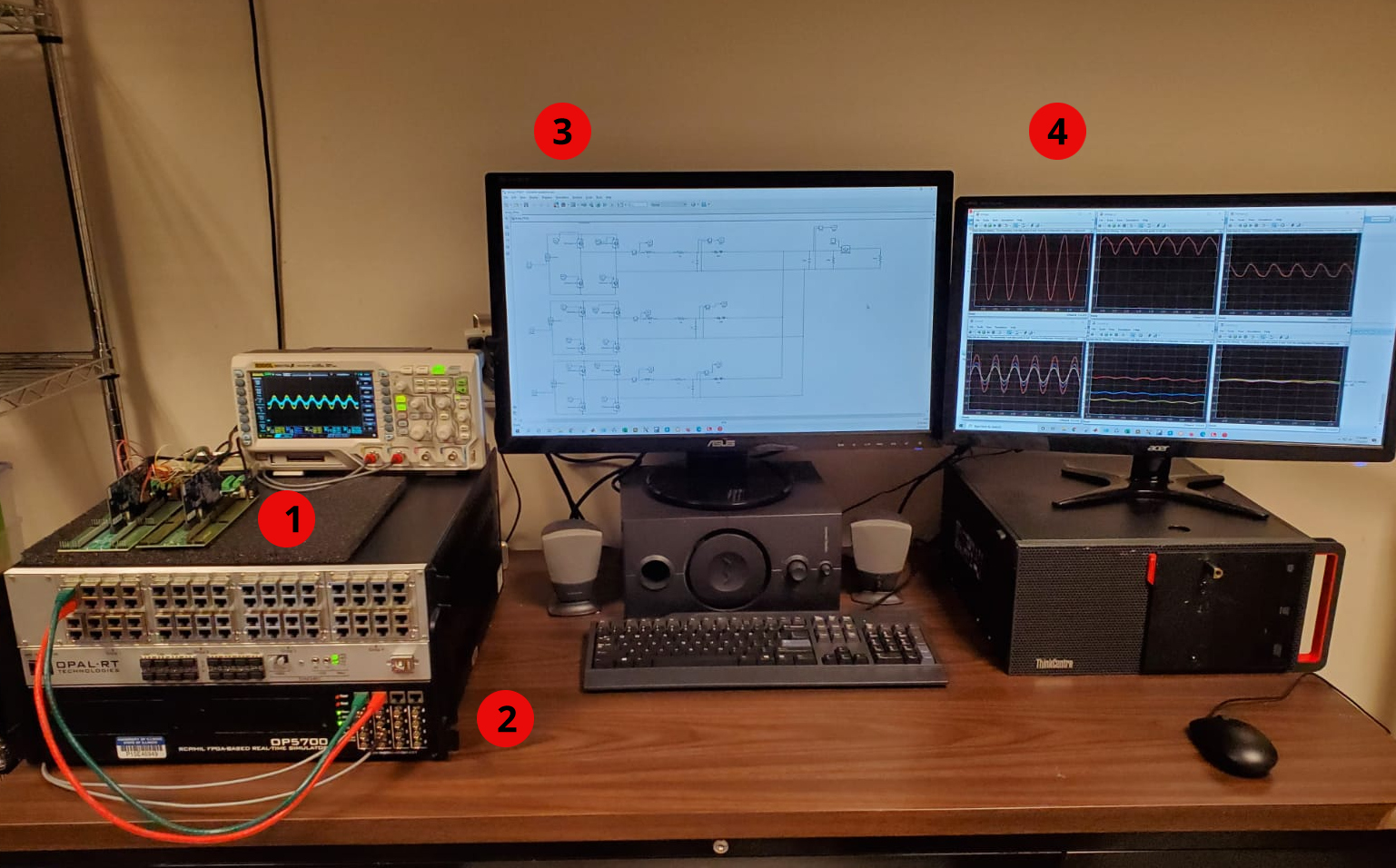}
    \caption{Experimental setup with (1) the controllers implemented on TMS320F28379D C2000 MCUs, (2) The OP5700 PHIL real-time simulation device, (3) switched model inverter and loads, (4) and real-time measurement signals.}
    \label{fig: Setup}
  \end{center}
\end{figure}
\begin{figure*}[!t]
    \centering
    \subfloat[]{\includegraphics[width=0.33\linewidth]
    {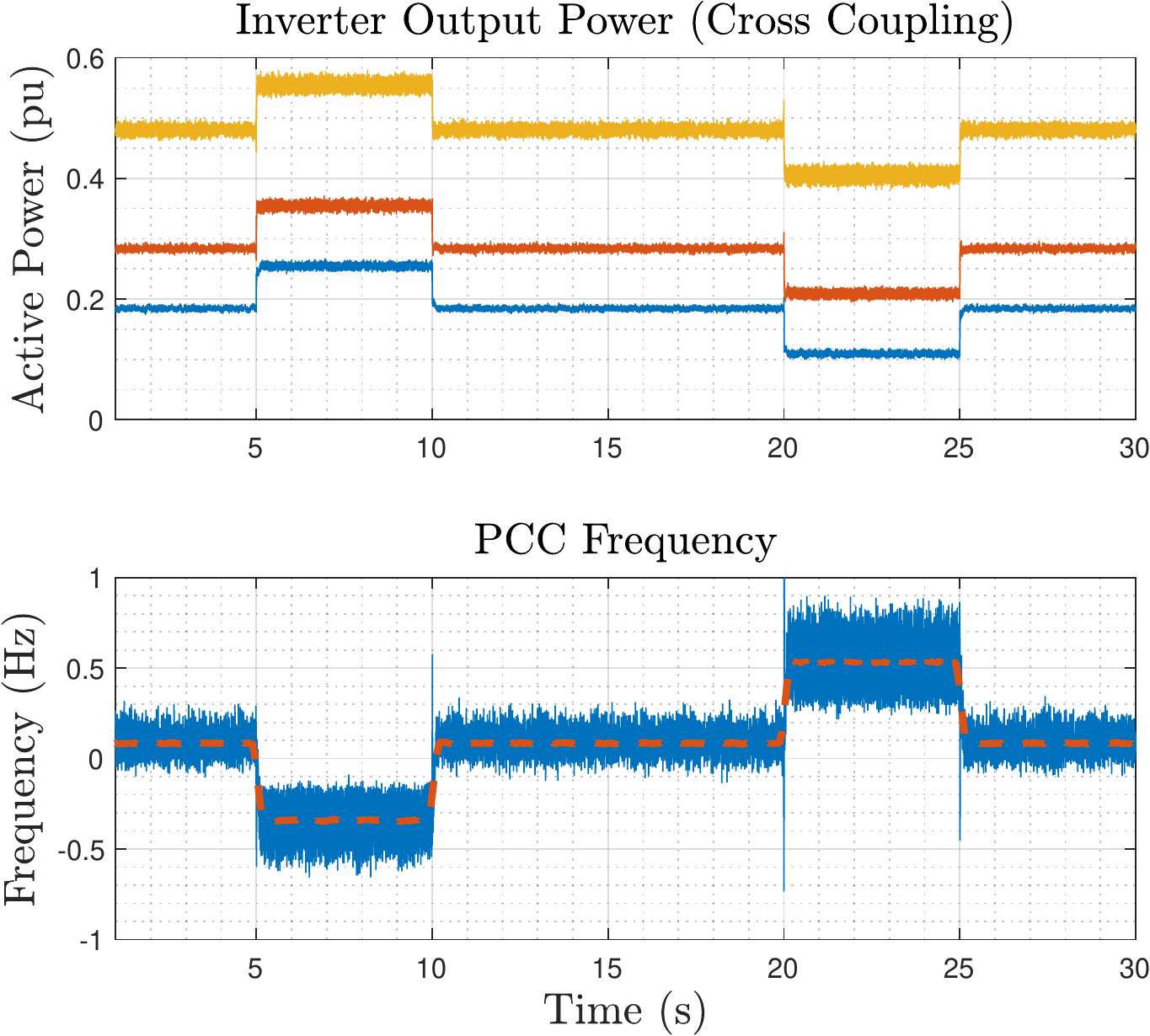}}
    \hfil
    \subfloat[]{\includegraphics[width=0.33\linewidth]
    {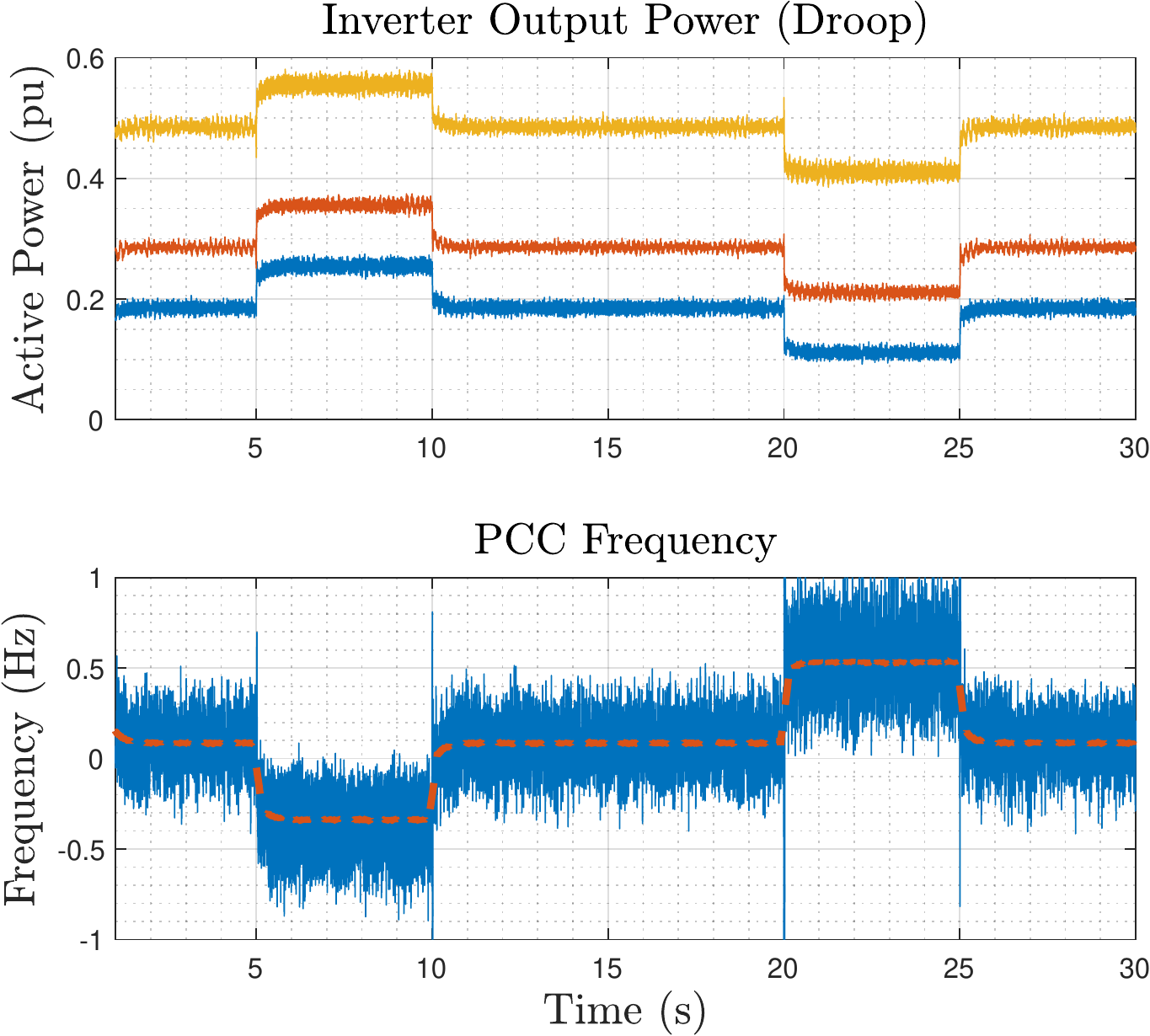}}
    \hfil
    \subfloat[]{\includegraphics[width=0.33\linewidth]
    {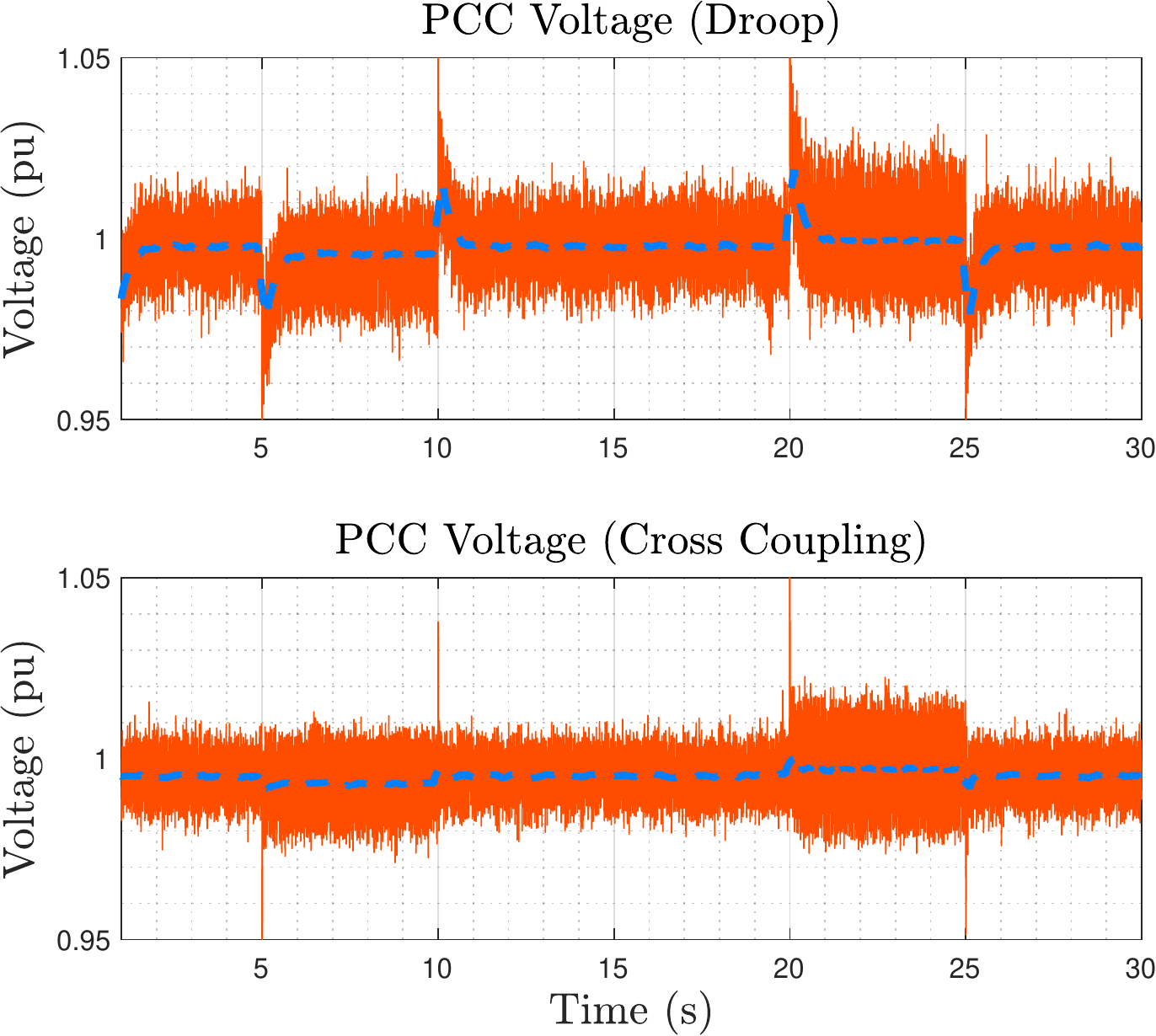}}
    \caption{(a) The output power of three parallel inverters with sharing ratios $\{0.5,0.3,0.2\}$pu. The transient behaviour of proposed control structure shows fast response to 20\% load change at PCC.(b) Three parallel droop-operated inverters with sharing ratios $\{0.5,0.3,0.2\}$pu. The transient behaviour is slower than proposed control.(c) The d component of PCC voltage $v^{d}$, shows near exact regulation of PCC voltage at 1pu for both droop and proposed control. However, the proposed control shows faster PCC voltage restoration in case of load change.}
    \label{fig: Droop_Compare}
\end{figure*}
\begin{figure*}
  \begin{center}
  \begin{minipage}[c]{0.35\textwidth}
    \subfloat[]{\includegraphics[width=1\linewidth]{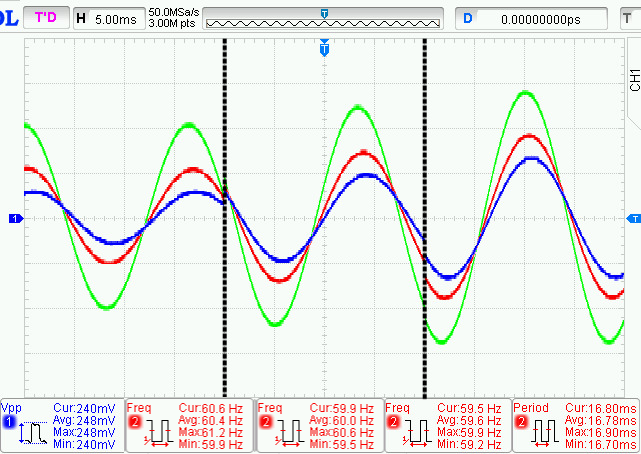}}
  \end{minipage}
  \begin{minipage}[c]{0.64\textwidth}
    \subfloat[]{\includegraphics[width=0.5\linewidth]{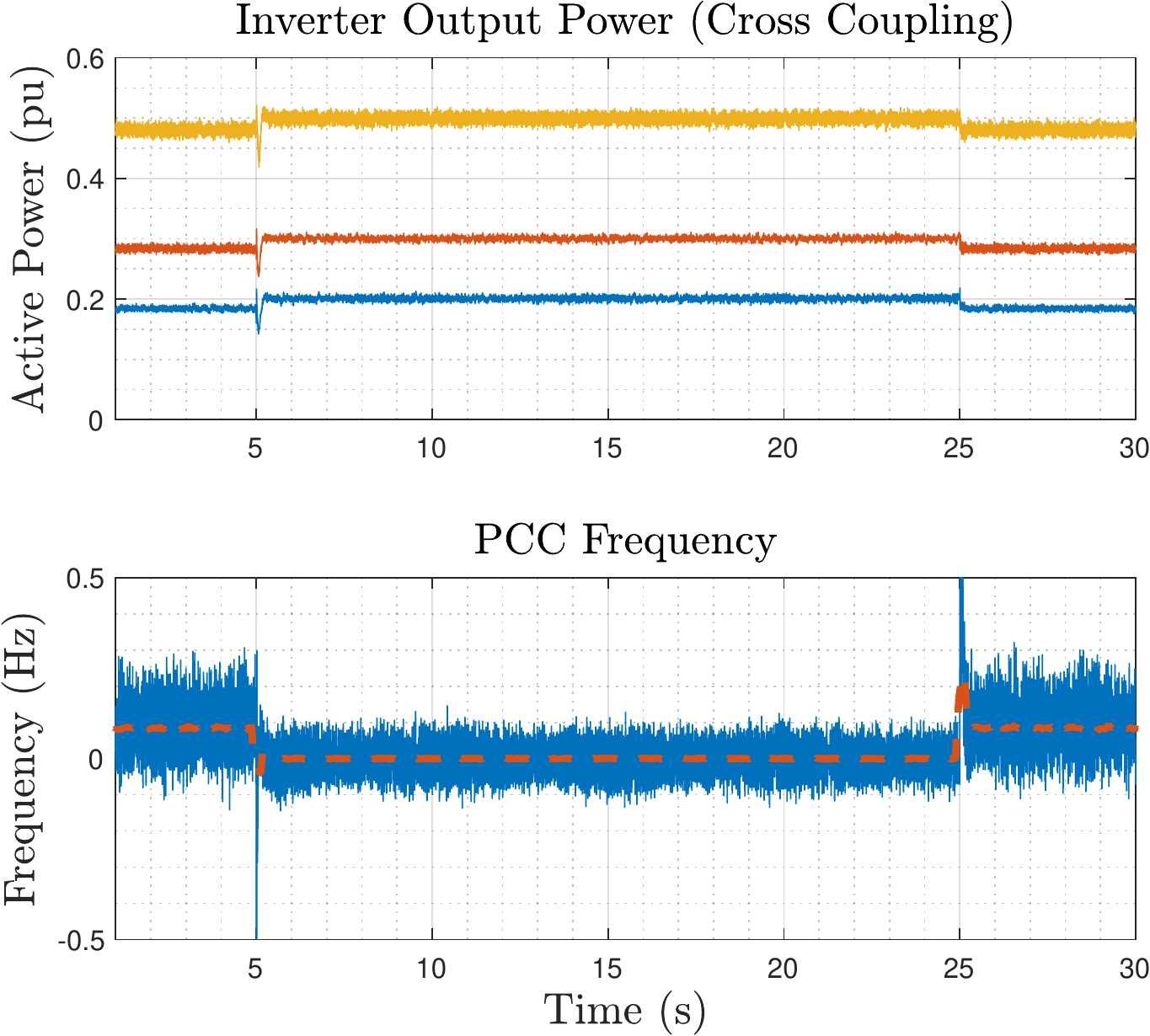}}
    \subfloat[]{\includegraphics[width=0.5\linewidth]{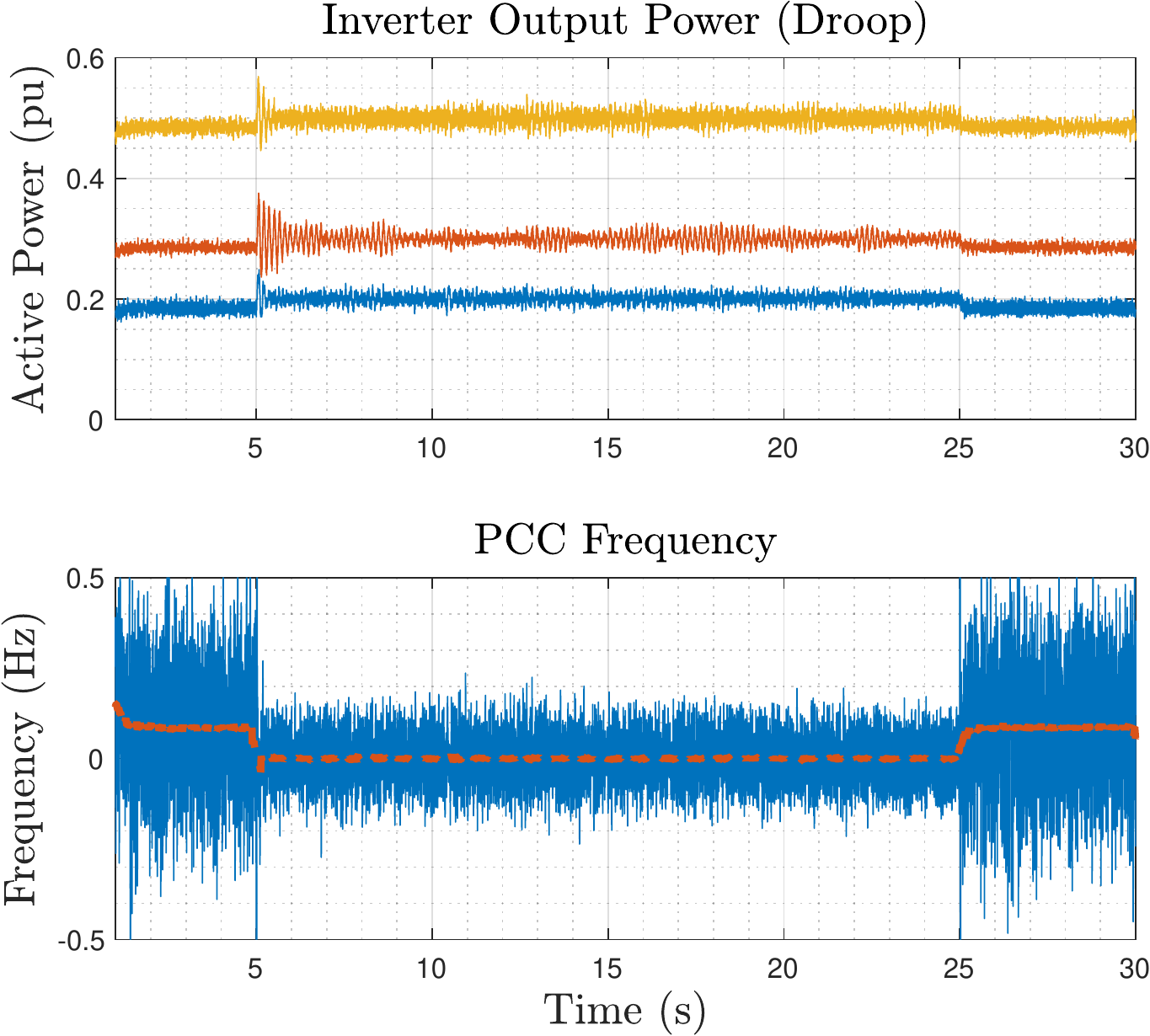}}
  \end{minipage}
    \caption{(a) The output current of inverters shown on The oscilloscope. The output current shows three distinct load conditions $\{0.8,1,1.2\}$pu, as depicted in Fig. \ref{fig: Droop_Compare}(a). The frequency measurement on the scope further validates the values obtained from sensor sampling. (b) Grid connection of proposed design at $t=5s$ followed by subsequent islanding at $t=25s$. The power setpoints for inverters are $\{0.5,0.3,0.2\}$pu, and the nominal load at PCC is 1pu. (c) The grid connection and islanding of droop design with the same setpoint and load condition as (b).}
    \label{fig: Grid_Con}
  \end{center}
\end{figure*}
The discretized controllers are implemented on Texas Instruments TMS320F28379D MCUs with discretization step $t_{s}=1/f_{sw}$, where $f_{sw}=20kHz$ is the switching frequency. The feedback and control signals from the real-time simulation are provided to embedded controllers through the OPAL-RT I/O interface.
\subsection*{System Performance} The controllers for each inverter system are designed using the structure and methodology described in section \uppercase\expandafter{\romannumeral 5}.B. However, to yield better harmonic performance, some of the gains are slightly changed from the calculated values. Moreover, to demonstrate robustness to modeling uncertainties, the controllers are designed based on the nominal value of $L_{i}=3\text{mH}$ for inverter inductor, $Z=0.1+j0.75\Omega$ for line impedance, and $40\mu\text{F}$ for filter capacitor. These nominal values are different from than real parametric values shown in the table\ref{table: param}.\\
In this paper, we consider three metrics for effective validation of the proposed controller. First is the precision of power sharing, second is the quality of voltage and frequency control, and finally, we consider the grid-tied operation of the controller.
\subsubsection*{Power sharing} In this part, we aim to evaluate the sharing performance of 3 parallel inverters operating in grid-forming mode. These inverters are completely operated in stand-alone mode with no information from other inverters. Additionally, the inverters are agnostic to load current $i_{load}$ and operate base on the provided nominal value $i_{0}$, set for each of them. The nominal current for each inverter forms a sharing ratio, as shown in the table\ref{table: param}.\\
Initially, the total load current is equivalent to 1pu that is equivalent to $i_{load}=100\text{A}$, and we set the respective set-points $i_{0}$, for each inverter as $\{20,30,50\}\text{A}$, satisfying the specified sharing ratio as given in parameter table. At nominal load, both the proposed design and droop attain the required power-sharing while maintaining both PCC voltage and frequency close to nominal values (see Fig. \ref{fig: Droop_Compare}(a)(b)(c)).\\ However, our proposed design achieves higher bandwidth and faster response to deviation in load current. We assess the transient behavior of the system by sudden step change of $\pm$20\% in load at $t=5s$ and $t=20s$. This deviation from nominal current allows us to analyze the inherent droop-like behavior of our design. From the experimental data, we validated the corresponding droop factor our design by observing approximately 0.4Hz deviation in nominal frequency for 0.2pu change in load. This is equivalent to the designed frequency droop coefficient of 2$\pi$[Hz/pu for three inverter systems]. Additionally, the design with cross-coupling term and inherent droop achieve approximately twice faster response in tracking the change in load current.
\subsubsection*{Voltage and Frequency Regulation}
In grid-forming microgrids, the voltage and frequency are collectively regulated by a set of grid-forming inverters. We showed in Fig. \ref{fig: Droop_Compare} that both droop operated and our control design achieve near-perfect voltage regulation. However, the voltage restoration in case of load change is twice faster in our proposed design. Additionally, based on the experimental validation, our design achieves signal waveform with less noise content compared to droop-operated VSI; this is evident when comparing the PCC frequency in Fig. \ref{fig: Droop_Compare}(a)(b) and voltage in Fig. \ref{fig: Droop_Compare}(c). The oscilloscope output current measurement of three inverters is given in Fig. \ref{fig: Grid_Con}(a).
\subsubsection*{Grid-tied operation} One of the main benefits of HSI is the grid-tied operation of the design. This design, by nature, has the capability to operate in both grid-forming and grid-tied modes. The assessment is done by first operating the system of parallel microgrids in the grid-forming mode where regulation of frequency and voltage is done by inverters while maintaining their prescribed sharing ratio (see Fig. \ref{fig: Grid_Con}(b)(c)). Then we connect the microgrid to the main grid by closing the smart breaker at $t=5s$. The smart breaker makes sure that the grid and microgrid are in phase at the moment of connection; the breaker checks no further requirement. In grid-tied operation, the voltage and frequency are maintained by the grid. Our proposed design shows a fast and seamless transition from islanded mode to grid ties operation with minimal transient following the connection. The droop-operated design shows oscillation at the moment of connection, and transients die out at a slower rate than our proposed design.\\
Finally, both the traditional droop and our novel approach achieve the same performance in power injection to the grid; however, our design seems more resilient and stable in grid-tied operation.
\section{Conclusion}
This paper presented a novel approach to design a hybrid sourced inverter that achieves a controlled compromise between current reference tracking and voltage regulation. The state the art approach to both control structure and control synthesis internalizes a droop-like behavior into the system dynamics without the requirement to deal with slow and non-linear power variables. This makes it possible for a more throughout analysis of system stability and performance than the more heuristic approaches taken with droop designs. As our future work, we aim to include an advanced $H\infty$ method to design the high-order controllers with complete stability and performance guarantee.


%




\ifCLASSOPTIONcaptionsoff
  \newpage
\fi



\bibliographystyle{IEEEtran}
\bibliography{IEEEabrv,Droop.bib}

\begin{thebibliography}{10}
\providecommand{\url}[1]{#1}
\csname url@samestyle\endcsname
\providecommand{\newblock}{\relax}
\providecommand{\bibinfo}[2]{#2}
\providecommand{\BIBentrySTDinterwordspacing}{\spaceskip=0pt\relax}
\providecommand{\BIBentryALTinterwordstretchfactor}{4}
\providecommand{\BIBentryALTinterwordspacing}{\spaceskip=\fontdimen2\font plus
\BIBentryALTinterwordstretchfactor\fontdimen3\font minus
  \fontdimen4\font\relax}
\providecommand{\BIBforeignlanguage}[2]{{%
\expandafter\ifx\csname l@#1\endcsname\relax
\typeout{** WARNING: IEEEtran.bst: No hyphenation pattern has been}%
\typeout{** loaded for the language `#1'. Using the pattern for}%
\typeout{** the default language instead.}%
\else
\language=\csname l@#1\endcsname
\fi
#2}}
\providecommand{\BIBdecl}{\relax}
\BIBdecl

\bibitem{inman2013solar}
R.~H. Inman, H.~T. Pedro, and C.~F. Coimbra, ``Solar forecasting methods for
  renewable energy integration,'' \emph{Progress in energy and combustion
  science}, vol.~39, no.~6, pp. 535--576, 2013.

\bibitem{liu2015comparison}
J.~Liu, Y.~Miura, and T.~Ise, ``Comparison of dynamic characteristics between
  virtual synchronous generator and droop control in inverter-based distributed
  generators,'' \emph{IEEE Transactions on Power Electronics}, vol.~31, no.~5,
  pp. 3600--3611, 2015.

\bibitem{chandorkar1993control}
M.~C. Chandorkar, D.~M. Divan, and R.~Adapa, ``Control of parallel connected
  inverters in standalone ac supply systems,'' \emph{IEEE transactions on
  industry applications}, vol.~29, no.~1, pp. 136--143, 1993.

\bibitem{chandorkar1996decentralized}
M.~Chandorkar and D.~Divan, ``Decentralized operation of distributed ups
  systems,'' in \emph{Proceedings of International Conference on Power
  Electronics, Drives and Energy Systems for Industrial Growth}, vol.~1.\hskip
  1em plus 0.5em minus 0.4em\relax IEEE, 1996, pp. 565--571.

\bibitem{golsorkhi2014control}
M.~S. Golsorkhi and D.~D. Lu, ``A control method for inverter-based islanded
  microgrids based on vi droop characteristics,'' \emph{IEEE Transactions on
  power delivery}, vol.~30, no.~3, pp. 1196--1204, 2014.

\bibitem{guerrero2005output}
J.~M. Guerrero, L.~G. De~Vicu{\~n}a, J.~Matas, M.~Castilla, and J.~Miret,
  ``Output impedance design of parallel-connected ups inverters with wireless
  load-sharing control,'' \emph{IEEE Transactions on industrial electronics},
  vol.~52, no.~4, pp. 1126--1135, 2005.

\bibitem{vasquez2009adaptive}
J.~C. Vasquez, J.~M. Guerrero, A.~Luna, P.~Rodr{\'\i}guez, and R.~Teodorescu,
  ``Adaptive droop control applied to voltage-source inverters operating in
  grid-connected and islanded modes,'' \emph{IEEE transactions on industrial
  electronics}, vol.~56, no.~10, pp. 4088--4096, 2009.

\bibitem{de2007voltage}
K.~De~Brabandere, B.~Bolsens, J.~Van~den Keybus, A.~Woyte, J.~Driesen, and
  R.~Belmans, ``A voltage and frequency droop control method for parallel
  inverters,'' \emph{IEEE Transactions on power electronics}, vol.~22, no.~4,
  pp. 1107--1115, 2007.

\bibitem{irving2000analysis}
B.~T. Irving and M.~M. Jovanovic, ``Analysis, design, and performance
  evaluation of droop current-sharing method,'' in \emph{APEC 2000. Fifteenth
  Annual IEEE Applied Power Electronics Conference and Exposition (Cat. No.
  00CH37058)}, vol.~1.\hskip 1em plus 0.5em minus 0.4em\relax IEEE, 2000, pp.
  235--241.

\bibitem{fantino2020grid}
R.~A. Fantino, C.~A. Busada, and J.~A. Solsona, ``Grid impedance estimation by
  measuring only the current injected to the grid by a vsi with $ lcl $
  filter,'' \emph{IEEE Transactions on Industrial Electronics}, vol.~68, no.~3,
  pp. 1841--1850, 2020.

\bibitem{ciobotaru2007online}
M.~Ciobotaru, R.~Teodorescu, P.~Rodriguez, A.~Timbus, and F.~Blaabjerg,
  ``Online grid impedance estimation for single-phase grid-connected systems
  using pq variations,'' in \emph{2007 IEEE Power Electronics Specialists
  Conference}.\hskip 1em plus 0.5em minus 0.4em\relax IEEE, 2007, pp.
  2306--2312.

\bibitem{asiminoaei2005implementation}
L.~Asiminoaei, R.~Teodorescu, F.~Blaabjerg, and U.~Borup, ``Implementation and
  test of an online embedded grid impedance estimation technique for pv
  inverters,'' \emph{IEEE Transactions on Industrial Electronics}, vol.~52,
  no.~4, pp. 1136--1144, 2005.

\bibitem{skogestad2007multivariable}
S.~Skogestad and I.~Postlethwaite, \emph{Multivariable feedback control:
  analysis and design}.\hskip 1em plus 0.5em minus 0.4em\relax Citeseer, 2007,
  vol.~2.

\bibitem{yazdani2010voltage}
A.~Yazdani and R.~Iravani, \emph{Voltage-sourced converters in power systems:
  modeling, control, and applications}.\hskip 1em plus 0.5em minus 0.4em\relax
  John Wiley \& Sons, 2010.

\bibitem{golestan2012dynamics}
S.~Golestan, M.~Monfared, F.~D. Freijedo, and J.~M. Guerrero, ``Dynamics
  assessment of advanced single-phase pll structures,'' \emph{IEEE Transactions
  on Industrial Electronics}, vol.~60, no.~6, pp. 2167--2177, 2012.

\bibitem{rocabert2012control}
J.~Rocabert, A.~Luna, F.~Blaabjerg, and P.~Rodriguez, ``Control of power
  converters in ac microgrids,'' \emph{IEEE transactions on power electronics},
  vol.~27, no.~11, pp. 4734--4749, 2012.

\bibitem{ferreira2019dynamic}
R.~V. Ferreira, S.~M. Silva, H.~M. Antunes, and G.~Venkataramanan, ``Dynamic
  analysis of grid-connected droop-controlled converters and synchronverters,''
  \emph{Journal of Control, Automation and Electrical Systems}, vol.~30, no.~5,
  pp. 741--753, 2019.

\bibitem{zhang2009power}
L.~Zhang, L.~Harnefors, and H.-P. Nee, ``Power-synchronization control of
  grid-connected voltage-source converters,'' \emph{IEEE Transactions on Power
  systems}, vol.~25, no.~2, pp. 809--820, 2009.

\end{thebibliography}
\end{document}